\documentclass[12pt]{article}

\usepackage[latin9]{inputenc}
\usepackage{geometry}
\usepackage{color}
\usepackage{array}
\usepackage{float}
\usepackage{units}
\usepackage{multirow}
\usepackage{amsmath}
\usepackage{graphicx}
\usepackage{setspace}
\usepackage[authoryear,round]{natbib}
\makeatletter

\providecommand{\tabularnewline}{\\}

\usepackage{multirow}
\usepackage{tabularx}
\usepackage{multicol}
\usepackage{version}
\graphicspath{{"C:/Users/carlos.santos/OneDrive - NOVASBE/Gradual_learning_gardete_santos/FiguresandTables"}}

\makeatother

\begin{document}
\title{\vspace{-2cm}
No data? No problem! A Search-based Recommendation System with Cold
Starts\thanks{Nova School of Business and Economics, Universidade NOVA de Lisboa,
Campus de Carcavelos, 2775-405 Carcavelos, Portugal. This work was
funded by Fundação para a Ciência e a Tecnologia (UID/ECO/00124/2019,
UIDB/00124/2020 and Social Sciences DataLab, PINFRA/22209/2016), POR
Lisboa and POR Norte (Social Sciences DataLab, PINFRA/22209/2016).\protect \\
Correspondence: Gardete: pedro.gardete@novasbe.pt, Santos: carlos.santos@novasbe.pt.}\vspace{1cm}
}
\author{Pedro M. Gardete \ \ \ \ Carlos D. Santos\vspace{1cm}
}
\date{October 2020}

\maketitle
\begin{spacing}{1.15}
{\small{}Recommendation systems are essential ingredients in producing
matches between products and buyers. Despite their ubiquity, they
face two important challenges. First, they are data-intensive, a feature
that precludes sophisticated recommendations by some types of sellers,
including those selling durable goods. Second, they often focus on
estimating fixed evaluations of products by consumers while ignoring
state-dependent behaviors identified in the Marketing literature.}{\small\par}

{\small{}We propose a recommendation system based on consumer browsing
behaviors, which bypasses the ``cold start'' problem described above,
and takes into account the fact that consumers act as ``moving targets,''
behaving differently depending on the recommendations suggested to
them along their search journey. First, we recover the consumers'
search policy function via machine learning methods. Second, we include
that policy into the recommendation system's dynamic problem via a
Bellman equation framework.}{\small\par}

{\small{}When compared with the seller's own recommendations, our
system produces a profit increase of 33\%. Our counterfactual analyses
indicate that browsing history along with past recommendations feature
strong complementary effects in value creation. Moreover, managing
customer churn effectively is a big part of value creation, whereas
recommending alternatives in a forward-looking way produces moderate
effects.}{\small\par}
\end{spacing}

\thispagestyle{empty}

\newpage{}

\section{Introduction}

Recommendation systems have been used in several industries to match
products and buyers. Often relying on machine learning techniques,
researchers have suggested recommendation systems that can affect
customers' choices and predict their preferences (e.g. Amazon recommendations,
Netflix ratings). Despite their popularity, modern recommendation
technologies face two challenges. The first, the ``cold start''
problem, arises when the data initially required to inform the recommendation
system is unavailable. The problem arises in multiple contexts: Small
companies often lack the resources necessary to acquire and maintain
customer- and product-related data, not to mention maintain the required
technological infrastructure. Durable goods sellers may also find
it challenging to acquire rich customer data: In our context, the
case of a North American online used-car seller, customers usually
buy at most one good, and thus generate little to no purchase history
data. Growing privacy concerns also limit firms' access to rich consumer-level
data. For example, the General Data Protection Regulation (GDPR) rules
out ``use-bundling,'' forcing sellers to disclose each intended
use of the data to be collected when asking consumers' permissions,
which is widely viewed as constraining for firms (\citealp{martin2019data}).

The second challenge -- as stated in \citet{AdomavicusTuzhilin2005}
-- is that ``in its most common formulation, the recommendation problem
is reduced to the problem of estimating ratings for the items that
have not been seen by a user. (...) Once we can estimate ratings for
the yet unrated items, we can recommend to the user the item(s) with
the highest estimated rating(s).'' Despite its prevalence, this approach
ignores the findings from the consumer search literature surrounding
the marked path-dependent behaviors by consumers.

In this paper, we develop a recommendation system that tackles these
two challenges. The recommendation system is rooted on an estimated
consumer ``search/purchase'' policy function. In line with the literature
on consumer search in Marketing, our recommendation system incorporates
the fact that consumers are a ``moving target,'' changing their
beliefs and behaviors as they learn more about the products offered
by the seller. This is done by first estimating the consumer's dynamic
policy function and second, using the estimated policy function to
inform the dynamic problem of the recommendation system. In addition
to determining the ex-ante first-best recommendation policy, we conduct
a series of counterfactual scenarios in order to determine the characteristics
of the recommendation system that drive value.

Our approach relies on the Bellman framework to propose recommendations
at each step of the consumer journey. This is important because we
know, from the literature on consumer search, that consumer behavior
is complex and path-dependent. For example, the seminal paper characterizing
stylized sequential search (\citealp{weitzman1979}) has found counterpoints
by \citet{deLosSantos2012TestingModelsConsumerSearch} and \citet{Honka_Chintagunta2017},
both of which find evidence towards fixed-sample search behaviors.
\citet{bronnenberg2016zooming} document non-trivial consumer search
behaviors, such as the focus on only a few characteristics in the
attribute space, and state dependence while navigating alternatives.
\citet{Ke_Shen_VillasBoas2016} and \citet{ursu2020search} incorporate
the continuous-type aspect of consumer search, and \citet{GardeteAntill2020}
account for piecemeal search for correlated characteristics of alternatives,
as observed in browsing data. 

This complex, path-dependent perspective is underrepresented in the
Computer Science literature on recommendation systems, whose main
object is to estimate consumers' expected utilities for different
alternatives. In that literature, recommendation systems are often
organized into two main types: Collaborative filtering and Content-based
filtering (see \citet{AdomavicusTuzhilin2005} for a review). Collaborative
filtering methods rely on the existence of users who exhibit overlaps
in terms of products bought (i.e., ``Buyers of this product also
bought...''). As for content-based filtering methods, these use the
characteristics of the items that users liked, bought, etc., and match
those behaviors to characteristics of other items available for sale.
For example, a user that listens to a particular music genre may be
recommended more songs of the same genre. Just like collaborative
filtering, content-based filtering also relies on broad availability
of user-data, such as past purchases. In practice, these methods rely
on a static view of consumer-product match values, and can perform
only when there exist significant amounts of data.

A third approach, relying on multi-armed bandits, assigns values to
multiple arms, i.e., the alternatives to be recommended. While these
methods are useful to address the cold start problem, they require
significant transformation in order to incorporate the dynamics of
consumer search. Finally, even recent advances in context-aware recommender
systems (e.g., \citealp{SongTekinSchaar2016}) tend to rely on static
contexts.\footnote{These methods also tend to optimize against click-through rates rather
than actual profitability. See \citet{KulkarniRodd2020} for a review
of context-aware recommendation systems. See also \citet{Agrawal2019}
for a review of MABs in the context of sequential decision-making.
In Marketing, multi-armed bandit models have been used for a variety
of tasks, including advertising placement (\citealp{SchwartzBradlowFader2017}).}

Taking advantage of a dataset containing clickstream data from a North
American online used car seller, we first estimate the consumers'
``search/purchase'' policy functions flexibly via machine learning
methods.\footnote{See \citet{Bajari_nekipelov_ryan_yang} for a review and application
of several machine learning methods for estimating demand models.} This procedure allows us to characterize the effect of product recommendations
and other customer actions on the subsequent probabilities of search
and conversion. To address the scalability issues associated with
the high number of alternatives involved, we rely on alternative-level
clustering (as in \citet{SongTekinSchaar2016}). In this case, ``the
recommendations are made at a cluster level instead of an item level.''
The consumer policy is estimated via multiple machine learning methods
and for several numbers of clusters. We analyze a number of out-of-sample
fit metrics in order to decide the optimal pair of estimation method
and number of clusters to be used.\footnote{An important question is the validity of the recovered policy function
when conducting counterfactual analyses. We take advantage of our
institutional setup, as discussed in detail in Section \ref{sec:ConsumerBehavior}.}

Once consumer search is characterized, we calculate the ex-ante performance
of the seller's recommendation system (i.e., the status quo), and
compare it with the performance of the ex-ante first-best system.
Our dynamic framework is able to anticipate complex consumer search
strategies. For example, our recommendation system may engage in active
learning by suggesting a potentially low-valued but highly informative
alternative, in order to transition the consumer to a state that makes
her more amenable to buying. Our model can also weigh promoting consumer
engagement against the likelihood of consumer churn.\footnote{See also the related work by \citet{Chen_Yao2017} and \citet{delossantoskoulayev2017}.
The latter ``propose a method of determining the ranking of search
results that maximizes consumers' click-through rates.'' In contrast
to this work, our model focuses on profitability rather than click-through
rates, and takes the stage of the consumer search journey into account
when recommending alternatives. } 

When compared with the status quo case, the first-best recommendation
system increases expected seller profits by 33\%. The system tends
to concentrate vehicle recommendations as the consumer search journey
progresses, being more likely to recommend vehicles from different
clusters near the beginning, and honing into a few or a single recommended
cluster down the line.

We conduct a series of counterfactual analyses in order to add interpretability
to the model's performance. First, despite the sizable increment in
profitability, we find that the status quo recommendation system does
extremely well in terms of maximizing profit based only on the last
alternative viewed by the consumer. In fact, our model is unable to
attain a statistically significantly better performance when constrained
to optimize solely on the last alternative browsed. This could make
it seem that previous browsing history and past recommendations, excluding
the alternative currently being viewed, are relatively much more important
for profitability. Our findings point otherwise. When basing the recommendation
system only on previous consumer and recommendation histories while
ignoring the last alternative viewed, profits increase by only 2.7\%.
We find that there is a significant complementarity between past history
and information on the current vehicle being viewed, which combined
account for the 33\% maximum increase in profitability. Additional
counterfactual analyses reveal that value creation depends less on
moving consumers towards high-margin alternatives, and more on 1)
managing consumer churn and 2) being forward-looking in terms of anticipating
the effect of product recommendations on the consumer journey as a
whole.

The next section describes the dataset. Section \ref{sec:ConsumerBehavior}
describes the estimation and identification procedure for the consumer
policy function. Section \ref{sec:Recommendation-System} describes
vehicle clustering and formalizes the recommendation system. Section
\ref{sec:Empirical-Results} presents the estimation results, and
Section \ref{sec:Counterfactual-Analyses} presents the findings from
the counterfactual analyses. Section \ref{sec:Conclusion} concludes.

\section{Data}

Our data was collected from a online platform that sells used cars,
Shift.com. This seller operates in multiple North American cities,
and allows buyers to book test-drives before they buy. The dataset
covers the period between February and September 2016. It includes
the full consumer clickstream data as well as the characteristics
of the vehicles on sale during the time period. The dataset is explained
in detail by \citet{GardeteAntill2020}. In contrast with their work,
we use all of the vehicle categories available for the analysis.

We observe a total of 71,143 users and 442,392 pageviews (Table \ref{tab:Descriptive-statistics-for}).
Each vehicle page was viewed an average of 130 times in our sample
and we denote significant variance: One vehicle exhibited a single
visit while another was visited more than 2,000 times.

\begin{figure}[H]
\caption{\textit{\emph{\label{fig:Nsearch_histogram} Histogram of Pageviews
per User}}}
\vspace{0.5cm}

\centering{}%
\begin{tabular}{c}
\includegraphics[width=0.8\columnwidth]{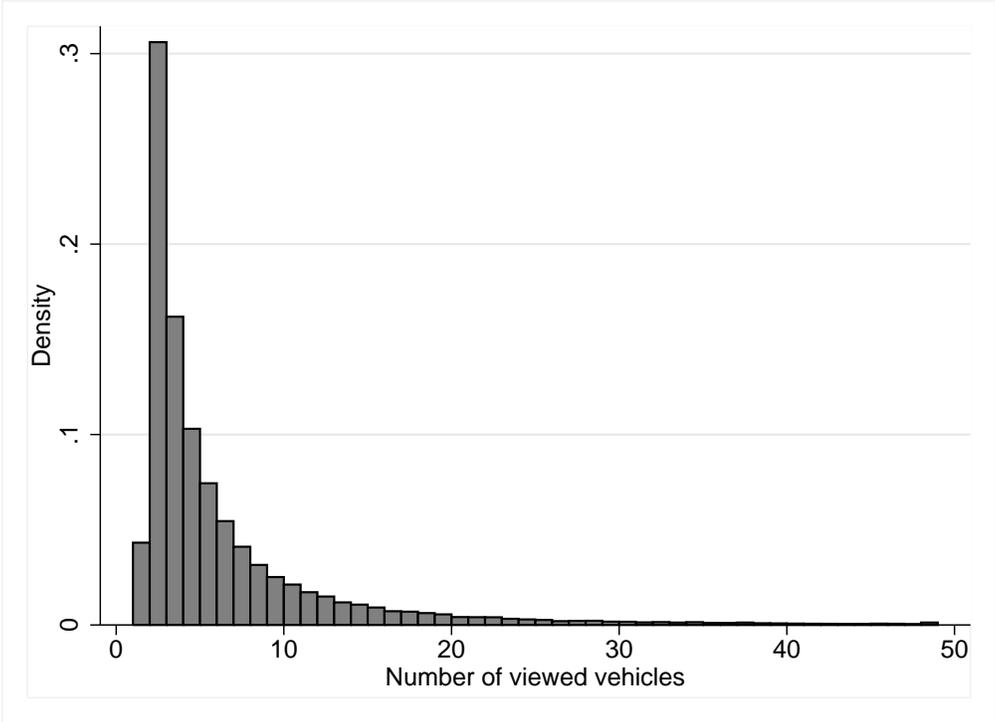}\tabularnewline
{\footnotesize{}}%
\begin{minipage}[t]{12cm}%
{\footnotesize{}Note: Distribution of pageviews per user, truncated
at the 99$^{th}$ percentile.}%
\end{minipage}\tabularnewline
\end{tabular}
\end{figure}

Consumer visits are tracked via IP address and cookies. A consumer
visit yields an average of 7 vehicle pageviews and 95\% of visitors
perform fewer than 22 searches (see Figure \ref{fig:Nsearch_histogram}).

\begin{table}[H]
\centering{}\caption{\textit{\emph{Descriptive Statistics for Browsing Data\label{tab:Descriptive-statistics-for}}}}
\vspace{0.5cm}
\begin{tabular}{lrrrrrr}
 & \multicolumn{1}{c}{} & \multicolumn{1}{c}{} & \multicolumn{1}{c}{} & \multicolumn{1}{c}{} & \multicolumn{1}{c}{} & \multicolumn{1}{c}{}\tabularnewline
\hline 
\hline 
Variables & Count & Mean & Std. Dev. & Min & Median & Max\tabularnewline
\hline 
Number of pageviews per user & \multicolumn{1}{r}{71,143} & \multicolumn{1}{r}{7.01} & \multicolumn{1}{r}{15.58} & \multicolumn{1}{r}{1} & \multicolumn{1}{r}{3} & \multicolumn{1}{r}{783}\tabularnewline
Conversion rate (\%) & \multicolumn{1}{r}{71,143} & \multicolumn{1}{r}{0.029} & \multicolumn{1}{r}{0.167} & \multicolumn{1}{r}{0} & \multicolumn{1}{r}{-} & \multicolumn{1}{r}{1}\tabularnewline
Number of visits per vehicle & 3,795 & 131.5 & 121.9 & 1 & 101 & 2,268\tabularnewline
Browsing time / pageview (min.) & \multicolumn{1}{r}{442,391} & \multicolumn{1}{r}{1.42} & \multicolumn{1}{r}{3.20} & \multicolumn{1}{r}{0} & \multicolumn{1}{r}{0.51} & \multicolumn{1}{r}{30}\tabularnewline
\hline 
\end{tabular}
\end{table}

Like \citet{GardeteAntill2020}, the conversion rate measures the
probability of booking a vehicle test-drive on the platform. We use
test-drives as proxies for purchase because 1) more than 90\% of consumers
who book test drives end up buying the vehicle; 2) because of point
1, there is little data available after the first test-drive (e.g.,
consumers seldom book a second one). We do not expect the relationship
between test-drive and purchases to be altered by changes to the recommendation
system, and so our counterfactual analyses should speak both to conversions
and to final purchases.

We also include vehicles' Carfax valuations at the time of the sample,
in order to construct potential vehicle selling margins. The main
goal with these data is to proxy for differential margins across vehicles
during counterfactual analyses. These data allow us to understand
the impact of having the recommendation system prioritize vehicles
with higher margins, for example.\footnote{In those analyses we assume 30\% margins, and exclude outliers below/above
the 5/95\% percentiles.}

Figure \ref{fig:Nsearch_histogram} presents the histogram of the
number of vehicle pageviews per user. Across their search journey,
most users (30\%) visit 2 vehicles. Moreover, behavior follows a long
tail: While most users (85\%) perform fewer than 10 searches, others
search beyond 30 (less than 3\%).

\begin{figure}[H]
\centering{}\caption{\textit{\emph{\label{fig:Recommendation-example.}Recommendation example}}}
\vspace{0.5cm}
\includegraphics[width=0.8\columnwidth]{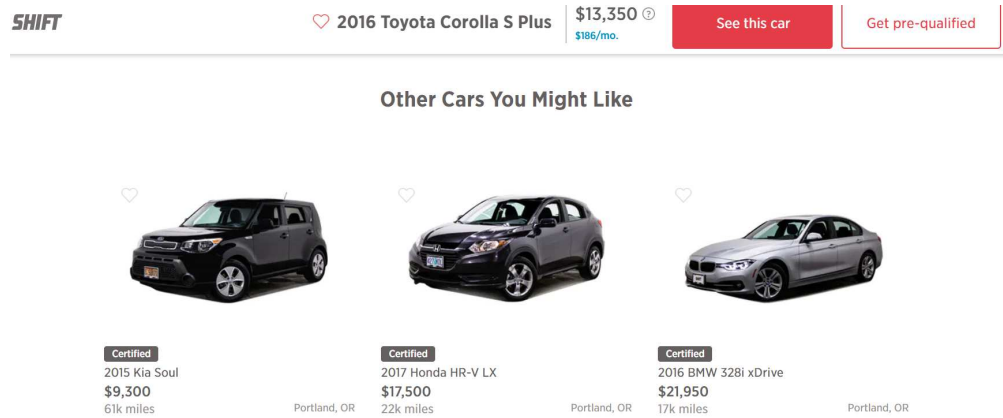}
\end{figure}

\textbf{\newpage Status Quo Recommendation System. }Every time a user
browses a vehicle, she is recommended 3 additional alternatives. Figure
\ref{fig:Recommendation-example.} shows the vehicle recommendations
to a user who viewed the vehicle profile page of a 2016 Toyota Corolla
S Plus.

The website's recommendation system suggests vehicles based on the
vehicle currently being viewed. These vehicles may be more or lesse
expensive than the focal vehicle. During the data collection period,
the seller's recommendation system was static: If a consumer visits
a vehicle and returns to it later, she will observe the same vehicle
recommendations. In other words, the status quo recommendation system
does not track consumer behavior in order to customize recommendations.

Figure \ref{tab:Probability-of-a-1} presents a probabilistic recommendation
matrix calculated from the status quo recommendation system. For this
exercise, vehicles were grouped into four distinct segments according
to cluster analysis (details are presented later, in Section \ref{subsec:Vehicle-Segments}).
Each row depicts a visit to a vehicle of one of the four clusters
and each column corresponds to a recommendation of a vehicle from
those same clusters.

{\small{}}
\begin{table}[H]
{\small{}\caption{\textit{\emph{\label{tab:Probability-of-a-1}Recommendation Probabilities
- 4 Clusters}}}
\vspace{0.5cm}
}{\small\par}
\centering{}%
\begin{tabular}{>{\centering}p{1cm}ccccc}
 &  & \multicolumn{4}{c}{}\tabularnewline
 &  & \multicolumn{4}{c}{Vehicle Recommendation}\tabularnewline
\cline{2-6} \cline{3-6} \cline{4-6} \cline{5-6} \cline{6-6} 
 & Cluster & 1 & 2 & 3 & 4\tabularnewline
\cline{2-6} \cline{3-6} \cline{4-6} \cline{5-6} \cline{6-6} 
 & 1 &  0.513  &  0.144  &  0.146  &  0.197 \tabularnewline
Vehicle & 2 & 0.145  &  0.433  &  0.193  & 0.229 \tabularnewline
Viewed & 3 &  0.191  &  0.252  &  0.362  &  0.195 \tabularnewline
 & 4 &  0.221  &  0.232  &  0.156  &  0.392 \tabularnewline
\cline{2-6} \cline{3-6} \cline{4-6} \cline{5-6} \cline{6-6} 
 & \multicolumn{5}{c}{{\footnotesize{}}%
\begin{minipage}[t]{8cm}%
{\footnotesize{}Note: Above, probability of recommending a vehicle
in a column, conditional on a consumer browsing a vehicle in a row.}%
\end{minipage}}\tabularnewline
\end{tabular}
\end{table}
{\small\par}

The main diagonal reveals that the current recommendation system is
not random: The algorithm is most likely to recommend vehicles that
have similar characteristics to the ones just viewed by consumers.\\\\
\textbf{Vehicle characteristics. }Our data comprises characteristics
of 4,140 vehicles, specifically, make, model, color, year, price,
mileage, number of owners, market valuation, body style, transmission,
drivetrain, and number of accidents.

\begin{table}[htbp]
\centering{}\caption{\textit{\emph{\label{tab:Desc_stats_vehicles}Descriptive Statistics
for Vehicle Data (metric variables).}}}
\begin{tabular}{llllll}
 & \multicolumn{1}{c}{} & \multicolumn{1}{c}{} & \multicolumn{1}{c}{} & \multicolumn{1}{c}{} & \multicolumn{1}{c}{}\tabularnewline
\hline 
\hline 
Variables  & Mean & Std. Dev. & Min & Median & Max\tabularnewline
\hline 
Year  & \multicolumn{1}{r}{2011} & \multicolumn{1}{r}{2.89} & \multicolumn{1}{r}{2001} & \multicolumn{1}{r}{2011} & \multicolumn{1}{r}{2016}\tabularnewline
Number of Owners  & \multicolumn{1}{r}{1.38} & \multicolumn{1}{r}{0.94} & \multicolumn{1}{r}{0} & \multicolumn{1}{r}{1} & \multicolumn{1}{r}{7}\tabularnewline
Number of Accidents  & \multicolumn{1}{r}{0.08} & \multicolumn{1}{r}{0.30} & \multicolumn{1}{r}{0} & \multicolumn{1}{r}{0} & \multicolumn{1}{r}{3}\tabularnewline
Mileage  & \multicolumn{1}{r}{100} & \multicolumn{1}{r}{27.15} & 83.49 & \multicolumn{1}{r}{132.57} & 220.74\tabularnewline
Price  & \multicolumn{1}{r}{100} & \multicolumn{1}{r}{10.1} & \multicolumn{1}{r}{86.07} & \multicolumn{1}{r}{97.37} & \multicolumn{1}{r}{223.37}\tabularnewline
Market Value  & \multicolumn{1}{r}{100} & \multicolumn{1}{r}{11.41} & \multicolumn{1}{r}{81.81} & \multicolumn{1}{r}{97.11} & \multicolumn{1}{r}{234.57}\tabularnewline
\hline 
N: & 4,140 &  &  &  & \tabularnewline
 &  &  &  &  & \tabularnewline
\multicolumn{6}{l}{{\footnotesize{}}%
\begin{minipage}[t]{12cm}%
{\footnotesize{}Note: Descriptive statistics for the vehicles available
on the seller's platform. Above, variables mileage, price and market
value are not comparable, as each is transformed via a different normalization.}%
\end{minipage}}\tabularnewline
\end{tabular}
\end{table}

The ordinal vehicle characteristics are presented in Table \ref{tab:Desc_stats_vehicles}.
The price and mileage variables have been scaled and mean-shifted
for confidentiality purposes. The largest vehicle class is of sedans
with front wheel drive and automatic transmission.

\section{Consumer Behavior\label{sec:ConsumerBehavior}}

\subsection{Policy Function}

The Marketing literature often uses dynamic programs to characterize
consumer search behaviors (\citet{Sun2006}, \citet{kim_albuquerque_bronnenberg2010},
\citet{YaoMela2011}, \citet{Seiler2013}, \citet{ursu2020search},
\citet{GardeteAntill2020}). The idea is that consumers procure information
strategically to find product matches. In turn, the recommendation
system incorporates the consumer's dynamic behavior into its own dynamic
optimization problem: It decides what to recommend to consumers at
each stage of their journey. We start out by defining the consumers'
policy function that maximizes their value:
\begin{equation}
Pr(\left.y_{it}=j\right|\Theta_{it}),\,j\in\{0..2J+1\}\label{eq:Pr_yit}
\end{equation}
Above, $y_{it}$ is consumer $i$'s action at time $t$ and $\Theta_{it}$
is the set of the consumer's current state variables. Set $\{0..2J+1\}$
includes all of the consumer's allowed actions at time $t$: Converting
immediately to one of the available vehicles ($J+1$ actions, including
the outside option), or deciding to search a specific vehicle ($J$
additional actions).

The consumer's policy function can be recovered flexibly from choice
data. Indeed, the goal of this first stage is to characterize consumer
behavior so it can be incorporated into the recommendation system's
optimization problem. When solving its own optimization problem, the
recommendation system will decide which vehicles to serve by incorporating
consumers' behaviors as described in \eqref{eq:Pr_yit}. Counterfactual
analyses rely on the consumer policy being recovered properly. We
discuss the identification of causal effects in the next section.

Once the consumer policy is estimated, it is used to calculate recommendations
across several counterfactual scenarios. Consumer policy \eqref{eq:Pr_yit}
is likely to remain valid across these analyses. Specifically, we
do not expect consumers to ascribe strategic behavior to the recommendation
system as we switch recommendation regimes. For one, such changes
are hard to identify from the consumer's point of view. Even if consumers
could indeed detect changes in recommendations, policy \eqref{eq:Pr_yit}
remains valid as long as they do not act strategically by attempting
to influence or game the recommendation system through their actions.
While such behaviors could occur in settings where consumers gain
a lot of experience with the platform's recommendation system (e.g.,
Amazon, YouTube), it is unlikely to arise in the context of the website
of a relatively small used car seller, where consumers often make
at most one purchase.\footnote{Note also that this assumption is used extensively in the empirical
literature on consumer search in Marketing and Economics and in the
Recommendation System literature in Computer Science.}

We include three groups of behavioral state variables while estimating
the consumer's policy function: The order of the interaction ($t$),
the vehicles browsed by the consumer up to $t$ ($H_{t}$), and the
history of vehicle recommendations ($R_{t}$). Moreover, we decompose
the vehicles browsed, $H_{t}$, into $H_{t}=\left\{ a_{t},A_{t}\right\} $,
where $a_{t}$ is the vehicle clicked by the user at time $t$ and
$A_{t}$ is the set of vehicles searched before time $t$.\footnote{This separation allows us to easily consider the case when the recommendation
system reacts only to the vehicle currently being viewed $(a_{t})$
rather than the full search history $H_{t}=a_{t}\cup A_{t}$. } The state space for consumer $i$ can be written as $\Theta_{it}=\left\{ t_{i},a_{it},A_{it},R_{it}\right\} $.
This space is used for policy estimation as well as for the purposes
of the recommendation system. We discuss the state space and its transitions
in more detail in Section \ref{sec:Recommendation-System}.

\subsection{Identification\label{sec:Identification}}

\begin{figure}[H]
\caption{\textit{\emph{\label{fig:Identification-of-the}Identification of
Causal Effects of Recommendations}}}
\vspace{0.5cm}

\centering{}{\small{}\includegraphics[width=0.8\columnwidth]{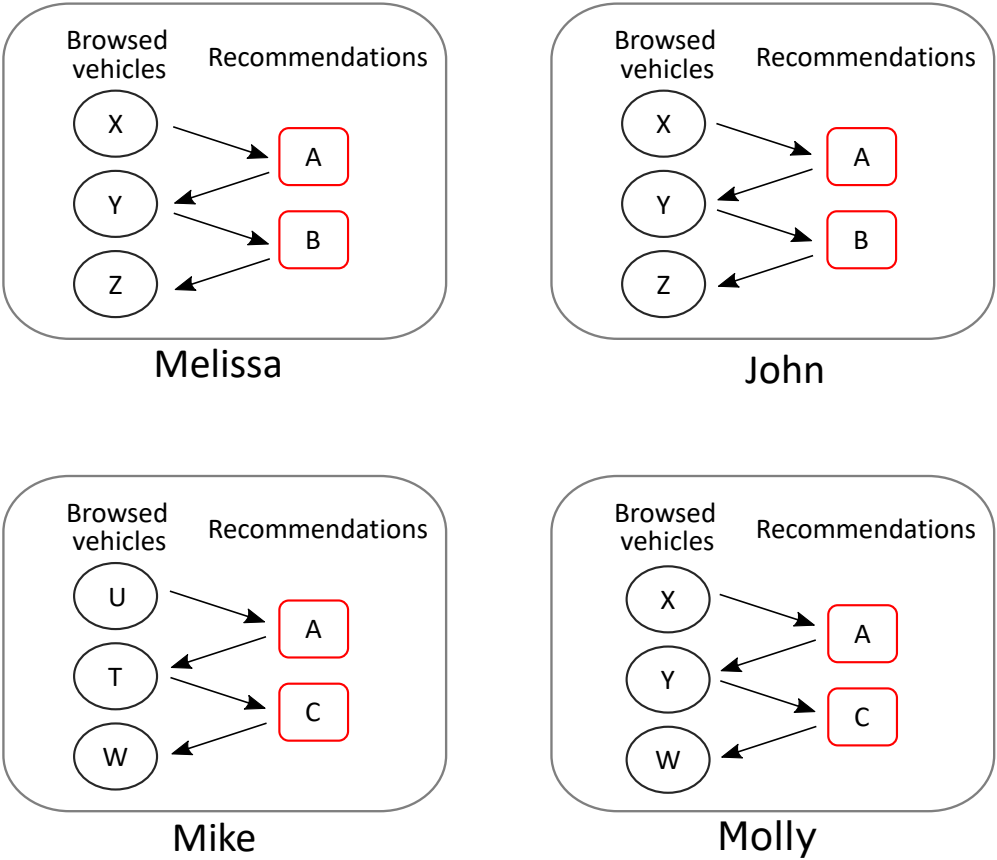}}{\small\par}
\end{figure}

It is worth considering how different recommendations interact with
policy \eqref{eq:Pr_yit}. Assigning a new set of recommendations
has two effects on consumers. First, it moves consumers onto new information
sets. For example, instead of learning about a ``green Mini,'' the
consumer may be recommended a ``tempting convertible Beetle.'' As
long as policy \eqref{eq:Pr_yit} is well recovered, we can use the
behavior of consumers in comparable states (i.e., those who were recommended
the convertible Beetle) to predict the counterfactual behavior of
the focal consumer. This strategy builds on i) policy \eqref{eq:Pr_yit}
being recovered flexibly and ii) the counterfactual analyses staying
within the ``support'' of the original recommendation system. We
address the first criterion by estimating policy \eqref{eq:Pr_yit}
via multiple machine learning methods across several numbers of vehicle
clusters, and selecting the winning pair via a number of out-of-sample
criteria. As for the support condition, we take advantage of the fact
that, as vehicles sell, the seller's recommendation system picks new
ones to recommend to customers. This natural rotation scheme allows
us to observe different vehicles being recommended to consumers in
the same state.\footnote{See also Tables \ref{tab:Probability-of-a-1} and \ref{tab:status-quo-rec8},
which show strictly positive recommendation probabilities across all
clusters, upon visiting each given vehicle class.} We illustrate identification through the example depicted in Figure
\ref{fig:Identification-of-the}. 

 In the example, four consumers browsed three vehicles and were exposed
to two product recommendations after each click. The leftmost consumers
(Mike and Melissa) are shown different recommendations after inspecting
their second vehicle. One could compare their subsequent behaviors
(visiting vehicle Z vs. W) to infer the effect of recommending vehicle
B rather than C. However, these consumers also feature different browsing
histories, a likely reflection of their preferences. Interpreting
differences in browsing behaviors by Mike and Melissa solely as the
consequence of the recommendations they were exposed to would not
produce valid causal effects of recommendations.

Melissa and John (topmost row of Figure \ref{fig:Identification-of-the})
feature similar browsing histories as a result of their similar preferences
for vehicles. However, in the example, both were exposed to the same
vehicles by the recommendation system. In this case the estimator
faces a crucial lack of variation. The examples discussed above illustrate
that, for proper identification of the effects of recommendations
(i.e., selection avoidance), we need consumers with \emph{similar}
preferences to be exposed to \emph{different} recommendations. 

We take advantage of the fact that vehicle recommendations change
as vehicles are sold in the platform. This explains the different
recommendation exposures by Mike and Molly in the example above, and
produces exogenous variation in the recommendations. We also benefit
from the fact that the status quo recommendation system only uses
the current vehicle being viewed as the basis for producing recommendations,
which allows us to control for it in the model directly, and ensures
that other sources of endogeneity are absent.\\\\
\textbf{Estimation.} The estimation procedure proceeds as follows:

1. Vehicles are grouped according to cluster analyses performed on
car characteristics. The analyses are performed once for each number
of vehicle clusters, from 3 to 10 clusters.

2. Consumer and vehicle data are organized according to the clusters
from step 1. For example, in the 3-cluster scenario, vehicle visits
are organized according to vehicle-cluster membership: A visit to
vehicles 1 and 2, both belonging to cluster 3, are re-coded as two
visits to vehicles of cluster 3.

3. Flexible supervised learning methods are used to estimate the consumer
policy \eqref{eq:Pr_yit}, once for each number of clusters (3 through
10; a total of $3\times8\text{=24}$ estimations). We implement decision
trees (random forests and boosting), and multinomial logit methods.\footnote{We also estimated a support vector machine model, but discarded it
due to the lack of fit and overall heavy computational burden.}

4. The estimation method is selected together with the number of clusters,
based on the cluster analysis silhouette values and out-of-sample
model fit metrics.

\section{Recommendation System\label{sec:Recommendation-System}}

\subsection{State Space: Vehicle Segments\label{subsec:Vehicle-Segments}}

As referred in the previous section, the recommendation system suggestions
rely on a dynamic program. We employ standard dynamic-programming
techniques to keep the size of the state space manageable. First,
we aggregate the 4,000 vehicles in our dataset into to clusters. The
goal is to produce near-homogeneous groups of vehicles to make the
problem estimable and more interpretable. As explained before, we
do not set the number of vehicle clusters a priori. Instead, we conduct
cluster analyses for multiple numbers of clusters, and select the
preferred number only after having estimated consumer behavior.

To overcome the local minimum problem of cluster analysis, we start
with a simple hierarchical solution (Ward's method) and use the resulting
allocation as the starting values for k-means clustering. This strategy
is likely to produce better allocations than using random initial
seeds. Vehicles are clustered by normalized values of their characteristics,
namely, body style, transmission type, drivetrain type, number of
accidents, number of owners, price, mileage, and markup. Categorical
variables are weighted by the number of different categories.\footnote{We employed two normalization alternatives for the continuous variables:
A min-max normalization and a quantile normalization. The advantage
of the quantile normalization is that it is invariant to order-preserving
transformations. Nonetheless, our results suggest that the min-max
normalization of the log-transformed variables, which we have opted
for, performs the best. Results are available from the authors.} 

\begin{figure}[H]
\centering{}{\small{}\caption{\textit{\emph{\label{fig:Optimal-Number-of}Silhouette Values by Number
of Clusters}}}
\includegraphics[width=0.8\columnwidth]{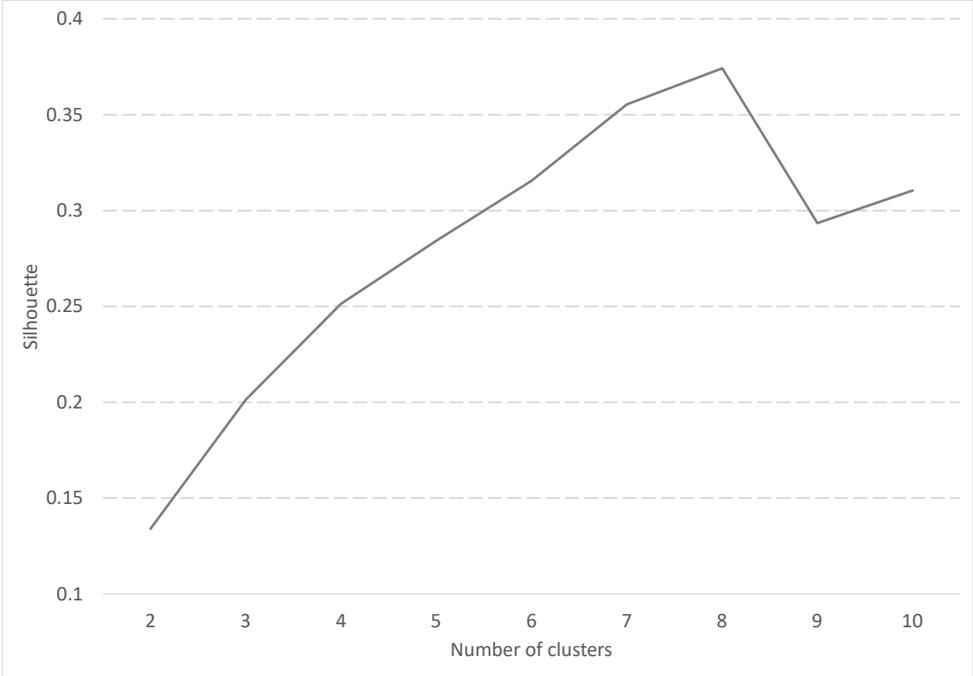}}
\end{figure}

We conduct cluster analyses while varying the number of clusters from
3 to 10. Figure \ref{fig:Optimal-Number-of} reports the silhouette
value for each of these analyses. The silhouette value is often used
in machine learning applications to select the optimal number of clusters.
In our case, the silhouette value is maximized at 8 clusters. 

{\small{}}
{\small\par}

We present some characteristics of the vehicle clusters in Table \ref{Tab: cluster_centroids},
for the case of 8 clusters. Cluster 1 may be labeled as rear-wheel
drive, since all its vehicles have rear-wheel drivetrains. Similarly,
all vehicles in cluster 2 are front-wheel drive vehicles. Cluster
3 is solely composed of vehicles with manual transmission and cluster
4 captures vehicles for which we have not been able to determine the
drivetrain with certainty from the website (and neither would consumers,
since we observe the same information they have had access to). Clusters
5, 6, and 7 feature vehicles with CVT, AWD and 4WD transmissions,
respectively. Finally, all vehicles in cluster 8 feature at least
one accident. In addition to the features we highlight here, the remaining
vehicle characteristics also vary across clusters, as can be verified
in Table \ref{Tab: cluster_centroids}.

\begin{table}[H]
\centering{}\caption{\textit{\emph{\label{Tab: cluster_centroids}}}Cluster Centroids -
Relative Frequencies}
\vspace{0.5cm}
\begin{tabular}{rrrr>{\raggedleft}m{1cm}rrrrr}
Cluster/Segment & 1 & 2 & 3 & 4 & 5 & 6 & 7 & 8 & Total\tabularnewline
\hline 
\multicolumn{1}{l}{Bodystyle} &  &  &  & \multirow{1}{1cm}{} &  &  &  &  & \tabularnewline
Convertible & 15\% & 3\% & 10\% & 8\% & 1\% & 2\% & 1\% & 6\% & 6\%\tabularnewline
Coupe & 18\% & 4\% & 28\% & 12\% & 4\% & 7\% & 0\% & 10\% & 10\%\tabularnewline
Hatchback & 1\% & 20\% & 26\% & 14\% & 34\% & 3\% & 1\% & 15\% & 15\%\tabularnewline
SUV & 6\% & 17\% & 7\% & 13\% & 16\% & 59\% & 78\% & 23\% & 21\%\tabularnewline
Sedan & 52\% & 51\% & 25\% & 51\% & 41\% & 24\% & 0\% & 40\% & 42\%\tabularnewline
Truck & 7\% & 0\% & 2\% & 0\% & 0\% & 0\% & 19\% & 1\% & 2\%\tabularnewline
Van & 0\% & 4\% & 0\% & 1\% & 0\% & 1\% & 0\% & 2\% & 1\%\tabularnewline
Wagon & 0\% & 2\% & 2\% & 2\% & 4\% & 4\% & 1\% & 2\% & 2\%\tabularnewline
\hline 
\multicolumn{1}{l}{Transmission} &  &  &  &  &  &  &  &  & \tabularnewline
Automatic & 100\% & 100\% & 0\% & 100\% & 0\% & 100\% & 100\% & 79\% & 78\%\tabularnewline
CVT & 0\% & 0\% & 0\% & 0\% & 100\% & 0\% & 0\% & 11\% & 11\%\tabularnewline
Manual & 0\% & 0\% & 100\% & 0\% & 0\% & 0\% & 0\% & 11\% & 12\%\tabularnewline
\hline 
\multicolumn{1}{l}{Drivetrain} &  &  &  &  &  &  &  &  & \tabularnewline
Rear WD & 100\% & 0\% & 31\% & 0\% & 0\% & 0\% & 0\% & 18\% & 22\%\tabularnewline
AWD & 0\% & 0\% & 11\% & 0\% & 19\% & 100\% & 0\% & 17\% & 14\%\tabularnewline
4WD & 0\% & 0\% & 4\% & 0\% & 1\% & 0\% & 100\% & 8\% & 5\%\tabularnewline
Front WD & 0\% & 100\% & 35\% & 0\% & 64\% & 0\% & 0\% & 41\% & 44\%\tabularnewline
NA & 0\% & 0\% & 18\% & 100\% & 16\% & 0\% & 0\% & 17\% & 14\%\tabularnewline
\hline 
\multicolumn{1}{l}{Number of accidents} &  &  &  &  &  &  &  &  & \tabularnewline
No accidents & 100\% & 100\% & 100\% & 100\% & 100\% & 100\% & 100\% & 0\% & 93\%\tabularnewline
At least one accident & 0\% & 0\% & 0\% & 0\% & 0\% & 0\% & 0\% & 100\% & 7\%\tabularnewline
\hline 
\multicolumn{1}{l}{Number of owners} &  &  &  &  &  &  &  &  & \tabularnewline
New car & 18\% & 17\% & 19\% & 18\% & 14\% & 16\% & 21\% & 0\% & 16\%\tabularnewline
1 owner & 28\% & 48\% & 42\% & 37\% & 60\% & 38\% & 47\% & 45\% & 43\%\tabularnewline
2 owners & 38\% & 28\% & 27\% & 34\% & 22\% & 35\% & 26\% & 42\% & 31\%\tabularnewline
3 or more owners & 15\% & 8\% & 12\% & 11\% & 4\% & 12\% & 6\% & 13\% & 10\%\tabularnewline
\hline 
\multicolumn{1}{l}{Normalized} &  &  &  &  &  &  &  &  & \tabularnewline
Log Transformed Price & 0.48 & 0.38 & 0.44 & 0.40 & 0.41 & 0.51 & 0.54 & 0.40 & 0.43\tabularnewline
Log Transformed Mileage & 0.86 & 0.85 & 0.83 & 0.85 & 0.83 & 0.86 & 0.85 & 0.88 & 0.85\tabularnewline
Log Transformed Margin & 0.42 & 0.34 & 0.39 & 0.38 & 0.36 & 0.45 & 0.47 & 0.36 & 0.38\tabularnewline
\multicolumn{1}{l}{Number of vehicles} & 725 & 1270 & 445 & 380 & 427 & 445 & 161 & 287 & 4,140\tabularnewline
\hline 
\hline 
\multicolumn{10}{l}{{\footnotesize{}}%
\begin{minipage}[t]{17cm}%
{\footnotesize{}Note: Cluster centroids for the eight cluster solution.
Centroids are the proportion (\%) of vehicles with a given characteristic,
except for the continuous variables (price, mileage and margin) where
the (logarithm) of the variables are min-max normalized.}%
\end{minipage}}\tabularnewline
\end{tabular}
\end{table}

\subsection{Dynamic Approach}

\begin{figure}[H]
\begin{centering}
{\small{}\caption{\textit{\emph{Illustration of State Variable Transitions\label{fig:state_space_sequence}}}}
}{\small\par}
\par\end{centering}
\centering{}\vspace{0.5cm}
{\small{}}%
\begin{tabular}{c}
{\small{}\includegraphics[width=0.7\paperwidth]{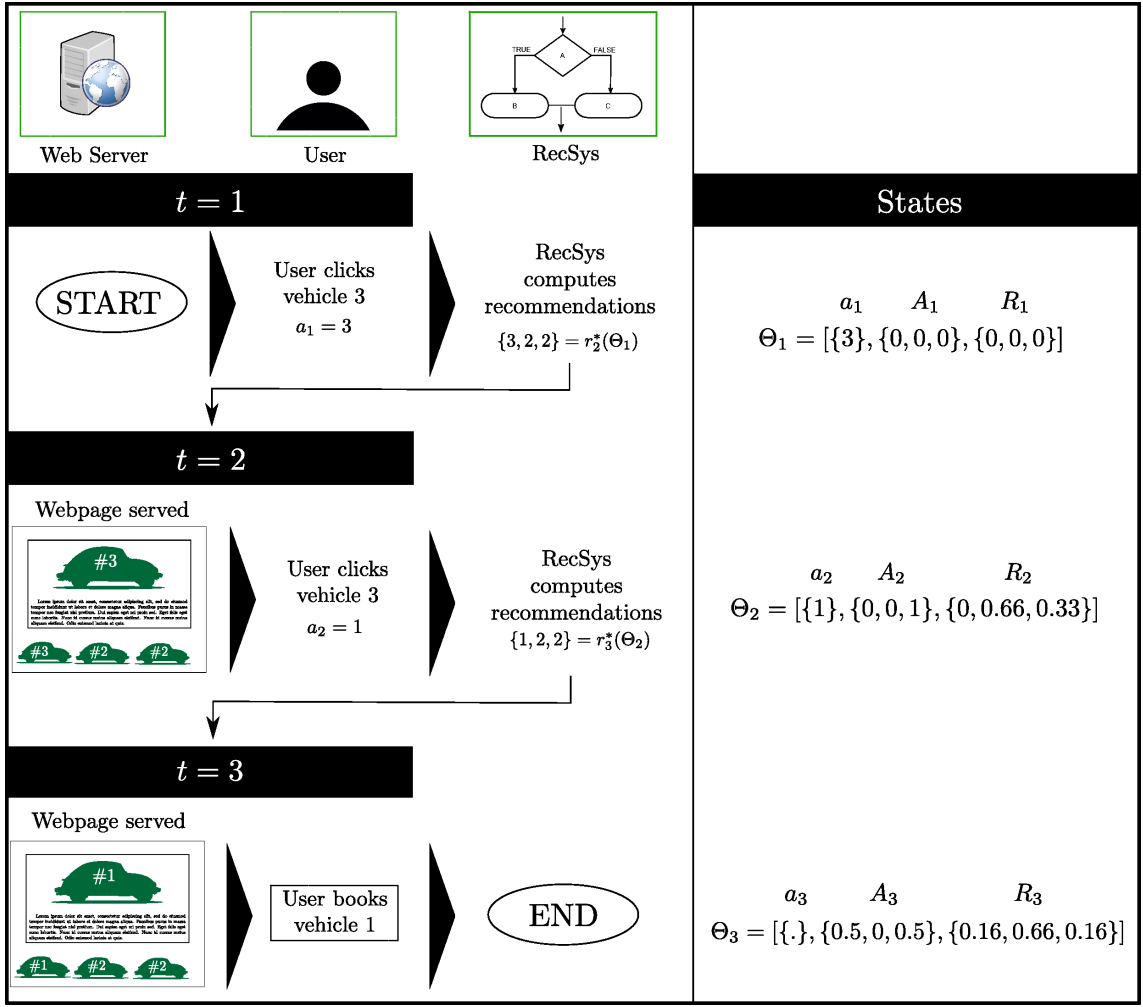}}\tabularnewline
\begin{minipage}[t]{14cm}%
{\footnotesize{}Note: In the example above there exist three groups
of vehicles that can be browsed by the consumer and recommended by
the recommendation system at each point in time. State variable $t$
omitted from the state variables for compactness.}%
\end{minipage}\tabularnewline
\end{tabular}{\small\par}
\end{figure}

The recommendation system suggests vehicles to consumers after each
click. The dynamic problem faced by the recommendation system can
be written through the following Bellman equation (subscript $i$
omitted):
\begin{align}
V\underset{\Theta_{t}}{\underbrace{\left(t,a_{t},A_{t},R_{t}\right)}} & =\max_{r_{t+1}}\,\pi\left(t,a_{t},A_{t},R_{t}\right)+E\left[\left.V\left(t+1,a_{t+1},A_{t+1},R_{t+1}\right)\right|a_{t},A_{t},R_{t},r_{t+1}\right]\label{eq:value_fn_fb-1}
\end{align}
The timing is a bit subtle and deserves discussion. Above, $a_{t}$
is the vehicle the consumer has just clicked in order to inspect it.
After this click, the recommendation system calculates the vehicles
to be recommended and the time period advances from $t$ to $t+1$.
Variables $A_{t}$ and $R_{t}$ contain the vehicles inspected and
recommended previously, respectively. Finally, $r_{t+1}$ is the set
of vehicles to be featured by the recommendation system at time $t+1$
- based on current period information. Much like online advertisement
impressions, recommendation $r_{t+1}$ is calculated immediately after
a consumer clicks on the link to a vehicle's webpage ($a_{t}$), but
before the new vehicle's page is rendered on the user's screen.\footnote{Recommendations can also be precalculated, and then selected depending
on the consumer's choice of vehicle to inspect.} The recommendation policy calculates three vehicle recommendations,
in line with the seller's website during the data collection period.
We denote the recommendation function as:
\begin{equation}
r_{t+1}=r^{*}\left(t,a_{t},A_{t},R_{t}\right)\label{eq:rec_function}
\end{equation}
 The result of this calculation is incorporated into the state variable
$R_{t+1}=f\left(R_{t},t,r_{t+1}\right)$, where $f\left(\cdot\right)$
is a transition function.\textbf{\\}\\
\textbf{State Variable Transitions. }The transition of variable $t$,
the order of the consumer's action, is straightforward. The consumer's
action $a_{t}$ originates from her policy, as characterized by expression
\eqref{eq:Pr_yit}. As for variables $A_{t}$ and $R_{t}$, these
keep track of the vehicles viewed by and recommended to the user in
the past. Specifically, the variables keep track of the relative frequencies
of vehicle clusters viewed and suggested in the past. For example,
if in the past a consumer has visited vehicles from cluster 1 three
times and has visited clusters 2 and 3 once each, the corresponding
state variable is equal to $A_{t}=\left\{ 0.6,0.2,0.2\right\} .$
This approach is based on the work by \citet{Santos2020}. The advantage
of keeping track of relative frequencies is that it reduces the size
of the state space by generating the same number of support points
as of a multinomial distribution. Following standard dynamic techniques
(state interpolation), we select grid values for the support of $A_{t}$
and $R_{t}$ and solve the value function through backward iteration.\footnote{We simulate back from 22 consumer searches, which in the dataset covers
approximately 95\% of the data.} We also take into account that the same values of $A_{t}$ and $R_{t}$
may mean different things to consumer behavior, depending on the stage
of the customer's journey. For this reason, we introduce the time
period $t$ as a state variable, making the problem non-stationary.
The introduction of state variable $t$ allows us to characterize
behavior differently at times, say, 3 and 5, even though $A_{3}=A_{5}$.
The added flexibility also applies, by construction, to all remaining
state variables.\\\textbf{}\\
\textbf{Example. }Figure \ref{fig:state_space_sequence} illustrates
the relationship between the consumer's search journey and the state
variable transitions. The consumer starts out by inspecting a vehicle
from group 3 and, based on this choice, the recommendation system
serves up recommendations $\left\{ 3,2,2\right\} $, i.e., one vehicle
from cluster 3 and two vehicles from cluster 2 are recommended.\footnote{Variables $A_{1}$ and $R_{1}$ are initialized at zero; these are
placeholder values for $t=1$, since there is no history yet, given
the cold start nature of the problem.} The web server then serves up the vehicle profile page along with
the vehicle recommendations. {\small{}}
{\small\par}

In period $t=2$, the customer decides to visit a vehicle from group
1. At this moment the recommendation system is called upon again to
decide which vehicles to feature. The state corresponding variables
are given by 
\begin{equation}
\Theta_{2}=\left\{ t,a_{2},A_{2},R_{2}\right\} =\left\{ 2,1,\left\{ 0,0,1\right\} ,\left\{ 0,0.66,0.33\right\} \right\} \label{eq:state2}
\end{equation}
Variables $t=2$ and $a_{2}$ are straightforward. Variable $A_{2}$
comprises the relative frequencies of vehicle groups shown to the
consumer in the past, in this case, 100\% of past visits have been
of vehicles in group 3. Similarly, $R_{2}$ stores the relative frequencies
of previously recommended vehicle groups: two vehicles of group 2
and one vehicle of group 3 have been recommended, giving rise to state
$R_{2}=\left\{ 0,0.66,0.33\right\} $. In the example, the recommendation
system calculates recommendations through function

\begin{equation}
r_{3}=r^{*}\left(\Theta_{2}\right)=\left\{ 1,2,2\right\} \label{eq:rec1}
\end{equation}
Once the web server serves the requested profile page and recommendations,
the consumer converts to a vehicle of group 1, thus terminating the
recommendation system's decision problem. \footnote{The intertwining dynamic problems faced by the consumers and the recommendation
system are presented in Appendix \ref{Appendix:Model}.}

\section{Empirical Results\label{sec:Empirical-Results}}

In order to estimate equation \eqref{eq:Pr_yit} in the most flexible
way, we implement three machine learning techniques: multinomial logit,
random forests, and boosting. Estimation is performed across methods
and numbers of clusters. Finally, we compare out-of-sample fit metrics
to assess which method/number-of-clusters combination to use for the
counterfactual analysis.

\subsection{Out-of-Sample Fit}

We start by comparing fit across the different specifications and
for different numbers of segments. We keep a random sample with 40\%
of individuals as a holdout, for which we construct our fit metrics.

The related literature contains a vast set of metrics to compare fit
for a fixed number of cluster segments (i.e. holding fixed the number
of categories that the dependent variable can take). The most common
fit metric in the literature is accuracy, i.e., the number of correctly
predicted cases divided by the total number of cases. However, by
using predicted classes instead of probabilities, accuracy serves
our purpose poorly. For example, in a model with two classes, a trained
prediction with 51\%/49\% probabilities produces the same accuracy
metric than a prediction with 99\%/1\% probabilities. This is a well-known
issue of the accuracy criterion. In order to incorporate predicted
\emph{probabilites} rather than predicted \emph{classes} into our
analysis, we include the log-loss criterion (equivalent to the log-likelihood
value: $\sum_{j}y_{j}.\ln\hat{p_{j}}$, where $j$ is a class) and
the difference between predicted and observed probabilities. A popular
choice for the latter is the use of the Helling distance metric given
by $\sum_{j}\left(y_{j}^{\frac{1}{2}}-\widehat{p}_{j}^{\frac{1}{2}}\right)^{2}$,
which we use.\footnote{Absolute differences in probabilities generate the same qualitative
implications for our analysis.}

Both of these metrics account for fit using predicted probabilities
rather than predicted classes. An important note, which we will revisit,
is that none of these metrics are expected to remain constant as the
number of classes/clusters changes. Therefore, the emphasis of this
first analysis is to determine which method appears to fit the out-of-sample
data best, across different numbers of clusters.

Table \ref{tab:Fit-metrics} exhibits fit metrics across models and
numbers of classes, from 3 to 10. Model performance is extremely similar
across methods and numbers of clusters, and no dominant method emerges.
For example, the multinomial logit model is more accurate than the
random forest method when 3 clusters are considered, but comes out
behind for 8 vehicle clusters. Also, when the number of clusters is
fixed, no model dominates another across criteria. The last column
of Table \ref{tab:Fit-metrics} presents a simple average of the different
criteria across methods. There we find that the random forest method
is always at least as good as the remaining ones across criteria,
when equal weights are used to average across clustering scenarios.
\begin{table}[H]
\caption{\label{tab:Fit-metrics}Fit metrics}
\vspace{0.5cm}

\centering{}{\small{}}%
\begin{tabular}{cccccccccc}
 & \multicolumn{8}{c}{} & \tabularnewline
\hline 
\hline 
 & \multicolumn{8}{c}{{\small{}Number of segments}} & \tabularnewline
 & {\small{}3} & {\small{}4} & {\small{}5} & {\small{}6} & {\small{}7} & {\small{}8} & {\small{}9} & {\small{}10} & {\small{}Simple Average}\tabularnewline
\hline 
 & \multicolumn{8}{c}{{\small{}Tree Ensemble (Random Forest)}} & \tabularnewline
\hline 
{\small{}Accuracy} & {\small{}47.1\%} & {\small{}34.2\%} & {\small{}33.3\%} & {\small{}28.6\%} & {\small{}29.4\%} & {\small{}29.7\%} & {\small{}22.1\%} & {\small{}22.5\%} & {\small{}30.86\%}\tabularnewline
{\small{}Log-Loss} & {\small{}0.361} & {\small{}0.445} & {\small{}0.483} & {\small{}0.522} & {\small{}0.550} & {\small{}0.571} & {\small{}0.615} & {\small{}0.628} & {\small{} 0.522 }\tabularnewline
{\small{}Prob. difference} & {\small{}0.122} & {\small{}0.113} & {\small{}0.097} & {\small{}0.086} & {\small{}0.077} & {\small{}0.069} & {\small{}0.065} & {\small{}0.059} & {\small{} 0.086 }\tabularnewline
\hline 
 & \multicolumn{8}{c}{{\small{}Tree Ensemble (Boosting)}} & \tabularnewline
\hline 
{\small{}Accuracy} & {\small{}46.8\%} & {\small{}33.8\%} & {\small{}33\%} & {\small{}30.2\%} & {\small{}30.0\%} & {\small{}29.1\%} & {\small{}22.2\%} & {\small{}22.1\%} & {\small{}30.9\%}\tabularnewline
{\small{}Log-Loss} & {\small{}0.362} & {\small{}0.446} & {\small{}0.484} & {\small{}0.523} & {\small{}0.55} & {\small{}0.57} & {\small{}0.615} & {\small{}0.627} & {\small{} 0.522 }\tabularnewline
{\small{}Prob. difference} & {\small{}0.122} & {\small{}0.113} & {\small{}0.097} & {\small{}0.086} & {\small{}0.077} & {\small{}0.069} & {\small{}0.065} & {\small{}0.059} & {\small{} 0.086 }\tabularnewline
\hline 
 & \multicolumn{8}{c}{{\small{}Multinomial Logit}} & \tabularnewline
\hline 
{\small{}Accuracy} & {\small{}47.6\%} & {\small{}34.5\%} & {\small{}33.1\%} & {\small{}30.4\%} & {\small{}30.2\%} & {\small{}29.5\%} & {\small{}21.5\%} & {\small{}21.4\%} & {\small{}31\%}\tabularnewline
{\small{}Log-Loss} & {\small{}0.365} & {\small{}0.453} & {\small{}0.492} & {\small{}0.53} & {\small{}0.559} & {\small{}0.579} & {\small{}0.623} & {\small{}0.635} & {\small{} 0.53 }\tabularnewline
{\small{}Prob. difference} & {\small{}0.123} & {\small{}0.115} & {\small{}0.098} & {\small{}0.087} & {\small{}0.078} & {\small{}0.07} & {\small{}0.066} & {\small{}0.06} & {\small{} 0.087 }\tabularnewline
\hline 
 &  &  &  &  &  &  &  &  & \tabularnewline
\multicolumn{10}{c}{{\footnotesize{}}%
\begin{minipage}[t]{17cm}%
{\footnotesize{}Note: All metrics are calculated in the houldout sample
(40\% of the sample not used for estimation). Accuracy is defined
as the number of correctly predicted cases divided by the number of
cases. Log-loss is equivalent to the log likelihood value. Probability
difference is the difference between the averaged probability predicted
by the model and the outcomes.}%
\end{minipage}{\small{} }}\tabularnewline
\end{tabular}{\small\par}
\end{table}

\begin{figure}[H]
\centering{}{\small{}\caption{\textit{\emph{\label{fig:fit_metrics}Fit Metrics for the Random Forest
Model: Lift, Pseudo-$R^{2}$, and Silhouette}}}
}\vspace{0.5cm}
{\small{}}%
\begin{tabular}{c}
{\small{}\includegraphics[width=0.8\columnwidth]{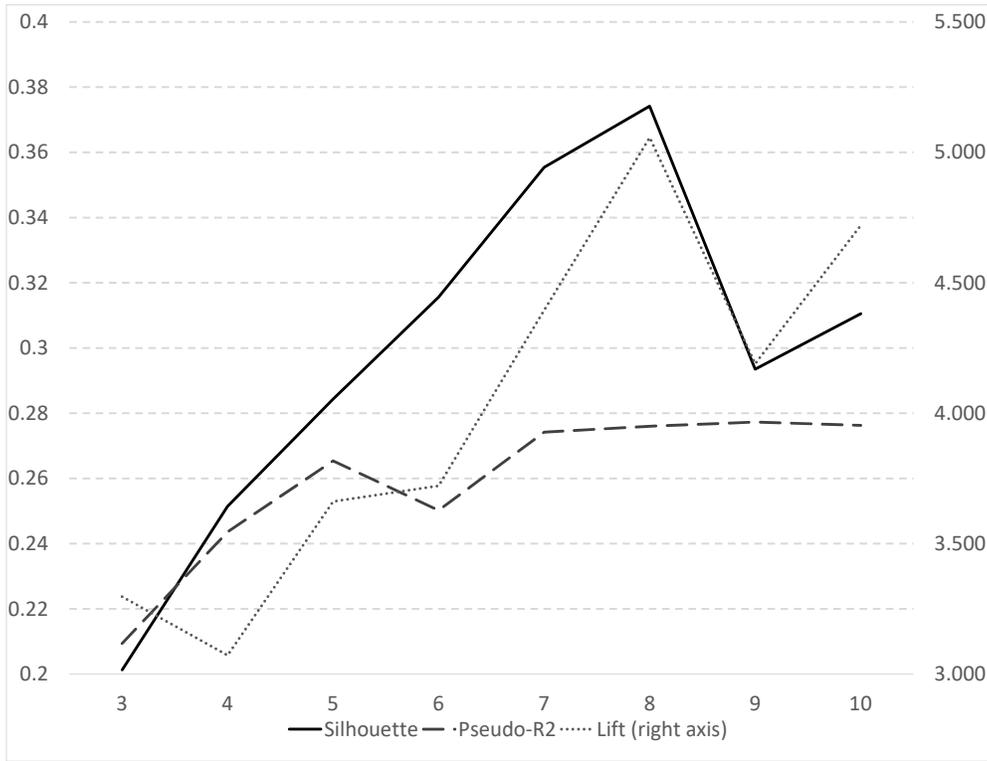}}\tabularnewline
{\footnotesize{}}%
\begin{minipage}[t]{12cm}%
{\footnotesize{}Note: Lift is defined as the number of correctly predicted
cases divided by the number of cases that would be predicted by a
random (uniform) model.}%
\end{minipage}\tabularnewline
\end{tabular}
\end{figure}

Due to their very nature, the metrics in Table \ref{tab:Fit-metrics}
are expected to be monotonic with the number of clusters. In order
to compare fit across numbers of clusters, we examine two additional
metrics: Lift and the Nagelkerke's pseudo-$R^{2}$. Figure \ref{fig:fit_metrics}
presents both metrics based on the random forest model, across numbers
of clusters. It also includes the silhouette metric from the cluster
analysis step, also reported in Figure \ref{fig:Optimal-Number-of}.

We find that the lift criterion is maximized at 8 clusters and the
pseudo-$R^{2}$ remains relatively stable between 7 and 10 clusters,
being highest at 9 clusters. The silhouette statistic from the cluster
analysis is maximized at 8 clusters. The last result is striking,
since the silhouette criterion is estimated independently from the
remaining metrics.{\small{}}
{\small\par}

Given the sharp decline of the lift and silhouette criteria at 9 clusters,
and the relative stability of the pseudo-$R^{2}$ metric around the
same region, we opt for the random forest model with 8 clusters from
here on. Given how close model performance is for any given number
of clusters, we expect that conducting the subsequent analyses with
other methods would yield very similar results.

\subsection{Model Predictions}

The status quo recommendation policy can be summarized by a matrix
that assigns probabilistic recommendations to each type of vehicle
viewed. Table \ref{tab:status-quo-rec8} displays the recovered status
quo recommendation policy at the selected number of clusters (eight).
\begin{table}[H]
\caption{Recommendation Policy - Status Quo\label{tab:status-quo-rec8}}
\vspace{1cm}

\centering{}%
\begin{tabular}{>{\centering}p{1cm}ccccccccc}
 &  & \multicolumn{8}{c}{Vehicle Recommendation}\tabularnewline
\cline{2-10} \cline{3-10} \cline{4-10} \cline{5-10} \cline{6-10} \cline{7-10} \cline{8-10} \cline{9-10} \cline{10-10} 
 & Cluster & 1 & 2 & 3 & 4 & 5 & 6 & 7 & 8\tabularnewline
\cline{2-10} \cline{3-10} \cline{4-10} \cline{5-10} \cline{6-10} \cline{7-10} \cline{8-10} \cline{9-10} \cline{10-10} 
 & 1 & \textbf{0.493} & \textbf{0.139} & \textbf{0.104} & 0.064 & 0.037 & 0.097 & 0.018 & 0.046\tabularnewline
 & 2 & \textbf{0.141} & \textbf{0.398} & 0.096 & \textbf{0.1} & 0.086 & 0.095 & 0.017 & 0.067\tabularnewline
 & 3 & \textbf{0.209} & \textbf{0.189} & \textbf{0.326} & 0.065 & 0.048 & 0.081 & 0.026 & 0.058\tabularnewline
Vehicle & 4 & \textbf{0.194} & \textbf{0.273} & \textbf{0.117} & \textbf{0.127} & 0.066 & \textbf{0.11} & 0.033 & 0.079\tabularnewline
Viewed & 5 & \textbf{0.133} & \textbf{0.317} & 0.089 & 0.078 & \textbf{0.201} & \textbf{0.105} & 0.033 & 0.046\tabularnewline
 & 6 & \textbf{0.236} & \textbf{0.184} & 0.071 & \textbf{0.105} & 0.04 & \textbf{0.259} & 0.063 & 0.042\tabularnewline
 & 7 & \textbf{0.172} & \textbf{0.194} & 0.099 & 0.081 & 0.041 & \textbf{0.212} & \textbf{0.165} & 0.036\tabularnewline
 & 8 & \textbf{0.216} & \textbf{0.256} & \textbf{0.125} & \textbf{0.106} & 0.079 & \textbf{0.136} & 0.03 & 0.053\tabularnewline
\cline{2-10} \cline{3-10} \cline{4-10} \cline{5-10} \cline{6-10} \cline{7-10} \cline{8-10} \cline{9-10} \cline{10-10} 
 & \multicolumn{9}{c}{{\footnotesize{}}%
\begin{minipage}[t]{12cm}%
{\footnotesize{}Note: Above, probability of recommending a vehicle
in a column, conditional on a consumer browsing a vehicle in a row.
Probabilities \textgreater{} 10\% are written in bold font.}%
\end{minipage}}\tabularnewline
\end{tabular}
\end{table}

The results are intuitive. First, the current algorithm privileges
recommending vehicles of the type being browsed, as visible in the
bold diagonal of vehicle clusters 1 through 7. In addition, vehicle
clusters 1 and 2 are also recommended more frequently, regardless
of the vehicle currently being viewed, which is unsurprising given
that they represent almost half of the vehicles in the sample. Vehicle
cluster 6 is also recommended across multiple cases.

Using the estimated policy function (equation \ref{eq:Pr_yit}), we
now compare its search, conversion, and recommendation predictions
with the analogue moments in the data. The results are presented in
Table \ref{tab:Probabilities}.

The model replicates the moments in the data extremely well, especially
in terms of recommendations and search decisions. The test-drive decision
(conversion) is not approximated as well for a number of segments.
\begin{table}[H]
\caption{\label{tab:Probabilities}Probabilities (Search, Recommendation, and
Conversion) - Data vs. Model Predictions}
\vspace{0.5cm}

\centering{}%
\begin{tabular}{ccccccccc}
 & \multicolumn{8}{c}{Probabilities}\tabularnewline
\cline{2-9} \cline{3-9} \cline{4-9} \cline{5-9} \cline{6-9} \cline{7-9} \cline{8-9} \cline{9-9} 
 & \multicolumn{2}{c}{Conversions} &  & \multicolumn{2}{c}{Recommendations} &  & \multicolumn{2}{c}{Search}\tabularnewline
\cline{2-3} \cline{3-3} \cline{5-6} \cline{6-6} \cline{8-9} \cline{9-9} 
 & Data & Simulated &  & Data & Simulated &  & Data & Simulated\tabularnewline
\hline 
Segment 1 & 0.53 & 0.30 &  & 25.93 & 24.71 &  & 20.59 & 16.00\tabularnewline
Segment 2 & 0.92 & 0.64 &  & 24.65 & 24.94 &  & 20.63 & 19.23\tabularnewline
Segment 3 & 0.21 & 0.30 &  & 12.87 & 12.78 &  & 10.91 & 10.24\tabularnewline
Segment 4 & 0.31 & 0.31 &  & 8.79 & 8.86 &  & 7.01 & 6.75\tabularnewline
Segment 5 & 0.30 & 0.34 &  & 6.67 & 6.92 &  & 5.74 & 6.78\tabularnewline
Segment 6 & 0.33 & 0.43 &  & 12.30 & 12.69 &  & 10.38 & 11.60\tabularnewline
Segment 7 & 0.06 & 0.06 &  & 3.28 & 3.53 &  & 2.76 & 3.87\tabularnewline
Segment 8 & 0.20 & 0.18 &  & 5.51 & 5.57 &  & 5.35 & 5.43\tabularnewline
\hline 
Outside option & 97.14 & 97.44 &  & - & - &  & - & -\tabularnewline
\hline 
 &  &  &  &  &  &  &  & \tabularnewline
\multicolumn{9}{l}{{\footnotesize{}}%
\begin{minipage}[t]{14cm}%
{\footnotesize{}Note: Probabilities for conversion, searches and recommendations
observed in the overall data (working and holdout samples) and predicted
by the model (simulation). Data truncated at 22 search actions, covering
approximately 95\% of users.}%
\end{minipage}}\tabularnewline
\end{tabular}
\end{table}
 This may be due to the relatively low conversion rate observed in
the data. We prefer not to `tweak' the model: Matching this moment
better could come at the cost of overfitting the data. As is common
in the Marketing literature, all subsequent counterfactual scenarios
are calculated within the estimated model, so that fit issues do not
affect the interpretation of the predictions.

\section{Counterfactual Analyses\label{sec:Counterfactual-Analyses}}

We now characterize the first-best recommendation system in depth,
and compare it with the status quo case. We then analyze the value-drivers
of the recommendation system in Section \ref{subsec:Rec.-System-Value}.

\subsection{First-Best Recommendation System}

The first-best recommendation system is obtained by solving the dynamic
problem of offering up recommendations. It incorporates the consumers'
decision probabilities and maximizes expected profit in a forward-looking
manner. Formally, for some consumer $i$ (subscript omitted), the
Bellman equation is given by:
\begin{align}
V\left(t,a_{t},A_{t},R_{t}\right) & =\max_{\varphi\left(\cdot\right)}\,\pi\left(t,a_{t},A_{t},R_{t}\right)+E\left[\left.V\left(t+1,a_{t+1},A_{t+1},R_{t+1}\right)\right|a_{t},A_{t},R_{t},\varphi\left(\cdot\right)\right]\label{eq:value_fn_fb}
\end{align}
where the state variables $\left\{ t,a_{t},A_{t},R_{t}\right\} $
characterize the probabilistic consumer search and conversion decisions,
as discussed in Section \ref{sec:ConsumerBehavior}. The payoff function
$\pi\left(\cdot\right)$ represents the expected instantaneous profit
for the seller: Here, the margins of the different vehicle clusters
are weighted by the respective instantaneous conditional purchase
probabilities. The expectation operator $E\left(\text{\ensuremath{\cdot}}\right)$
is a consequence of the uncertainty about the consumer's subsequent
action, conditional on the current state variables. The uncertainty
is characterized by the probability mass function
\begin{equation}
\text{Pr}(a_{t+1}|t,a_{t},A_{t},R_{t},\varphi\left(\cdot\right))=\text{Pr}(a_{t+1}|t,a_{t},A_{t},R_{t}\cup r_{t+1}),\label{eq:prob1}
\end{equation}
as induced by the estimated consumer's policy function.

The recommendation function is given by $\varphi\left(t,a_{t},A_{t},R_{t}\right)$:
It determines the set of three recommendations to be shown to user
$i$ immediately after she visits a vehicle profile. In between the
user's ``click'' and the impression of the new page, the recommendation
system analyzes the consumer's current state, given by $\left\{ t,a_{t},A_{t},R_{t}\right\} $,
and computes recommendation $\varphi\left(\cdot\right)$. Finally,
once the vehicle profile page is displayed to the consumer, state
variable $R_{t}$ transitions to $R_{t+1}=\left\{ R_{t}\cup\varphi\left(t+1,a_{t+1},A_{t+1},R_{t}\right)\right\} $,
where the ``union'' operator denotes the relative frequency update
of variable $R_{t}$ via the recommendations in $\varphi$. Variable
$A_{t+1}$ is updated via $a_{t}$ through an analogue procedure.

We calculate the first-best recommendation policy via backward induction.
In order to interpret the ex-ante first-best policy, we first calculate
the implied average transition matrix. In line with the status quo
recommendation matrix presented in Table \ref{tab:status-quo-rec8},
we consider only the last action taken by each consumer, and note
the probability of each vehicle recommendation offered by the first-best
policy. We present the results in Table \ref{tab:first-best-rec}.

\begin{table}[H]
\caption{Recommendation Policy - First-Best\label{tab:first-best-rec}}
\vspace{1cm}

\centering{}%
\begin{tabular}{>{\centering}p{1cm}ccccccccc}
 &  & \multicolumn{8}{c}{Vehicle Recommendation}\tabularnewline
\cline{2-10} \cline{3-10} \cline{4-10} \cline{5-10} \cline{6-10} \cline{7-10} \cline{8-10} \cline{9-10} \cline{10-10} 
 & Cluster & 1 & 2 & 3 & 4 & 5 & 6 & 7 & 8\tabularnewline
\cline{2-10} \cline{3-10} \cline{4-10} \cline{5-10} \cline{6-10} \cline{7-10} \cline{8-10} \cline{9-10} \cline{10-10} 
 & 1 &  0.021  & \textbf{ 0.506 } &  0.001  &  0.003  & \textbf{ 0.143 } & \textbf{ 0.122 } & \textbf{ 0.159 } &  0.044 \tabularnewline
 & 2 &  0.001  & \textbf{0.232 } &  0.035  &  0.003  & \textbf{ 0.465 } & \textbf{ 0.222 } & 0.038  &  0.004 \tabularnewline
 & 3 &  0.001  & \textbf{ 0.788 } &  0.038  &  0.003  &  0.010  & 0.078  & 0.072  &  0.009 \tabularnewline
Vehicle & 4 &  0.001  & \textbf{ 0.592 } &  0.009  & \textbf{0.1 } &  0.054  & \textbf{ 0.126 } & \textbf{ 0.117 } &  0.000 \tabularnewline
Viewed & 5 &  0.046  & \textbf{ 0.326 } & \textbf{ 0.152}  & \textbf{ 0.381 } &  0.021  &  0.062  &  0.008  &  0.004 \tabularnewline
 & 6 &  0.000  & \textbf{ 0.716 } & \textbf{ 0.123 } &  0.001  &  0.037  &  0.025  &  0.040  &  0.058 \tabularnewline
 & 7 &  0.025  & \textbf{ 0.779 } &  0.021  &  0.030  &  0.046  &  0.040  &  0.019  &  0.039 \tabularnewline
 & 8 &  0.059  & \textbf{ 0.627 } &  0.064  &  0.098  &  0.003  &  0.013  & \textbf{0.110 } &  0.026 \tabularnewline
\cline{2-10} \cline{3-10} \cline{4-10} \cline{5-10} \cline{6-10} \cline{7-10} \cline{8-10} \cline{9-10} \cline{10-10} 
 & \multicolumn{9}{c}{{\footnotesize{}}%
\begin{minipage}[t]{14cm}%
{\footnotesize{}Note: Above, probability of recommending a vehicle
in a column, conditional on a consumer browsing a vehicle in a row.
Remaining states are averaged out. Probabilities \textgreater{} 10\%
are written in bold font.}%
\end{minipage}}\tabularnewline
\end{tabular}
\end{table}

\begin{figure}[H]
\caption{Distribution of Vehicle Cluster Recommendations Over Time\label{fig:reg_3bg_dist_t}}
\vspace{0.5cm}

\centering{}%
\begin{tabular}{c}
\includegraphics[width=0.65\paperwidth]{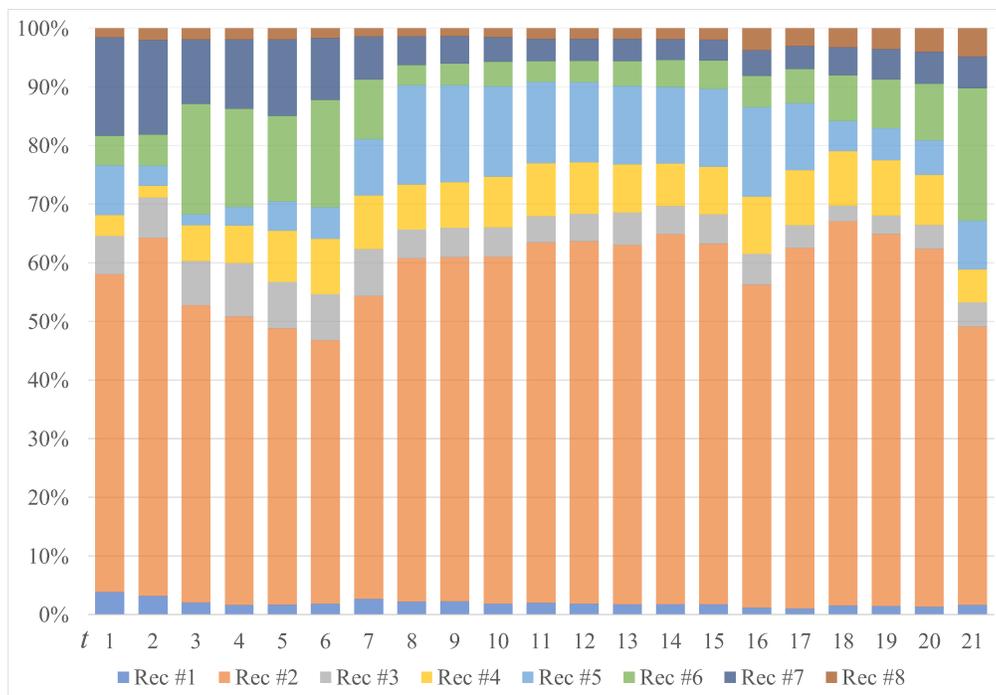}\tabularnewline
\begin{minipage}[t]{16cm}%
Note: Above, we depict 21 recommendations shown to consumers after
each click. In the model we consider a total of 22 ``clicks,'' the
first occurring before recommendations are made. In our data, 95\%
of consumers search vehicles at most 22 times.%
\end{minipage}\tabularnewline
\end{tabular}
\end{figure}

In comparison with the status quo policy (Table \ref{tab:status-quo-rec8}),
we find that the first-best policy 1) relies less on the last vehicle
viewed by the consumer and 2) is more likely to recommend vehicles
from cluster 2, independently of the previous action.

Table \ref{tab:first-best-rec} does not tell the full story, since
in reality it depends on a number of variables in addition to the
type of the last vehicle viewed. Figure \ref{fig:reg_3bg_dist_t}
shows the relative frequency of recommended vehicle clusters over
time.

The results in the figure above are averaged across states uniformly,
so that they can be interpreted independently of consumers' behaviors.
Figure \ref{fig:reg_3bg_dist_t} reveals that the first-best recommendation
system is far from stationary. For example, near the beginning of
the consumer journey, vehicles from clusters 7 are relatively more
likely to be recommended. By search number 8 however, cluster 7 is
seldom recommended, giving space to vehicles from clusters 2 and 5.
Near the end of the consumer journey, these clusters are substituted,
on average, by recommendations of vehicles from cluster 6.

\begin{figure}[H]
\caption{Distribution of Different Cluster Recommendations Over Time\label{fig:reg_3bg_var}}

\centering{}\includegraphics[width=0.7\paperwidth]{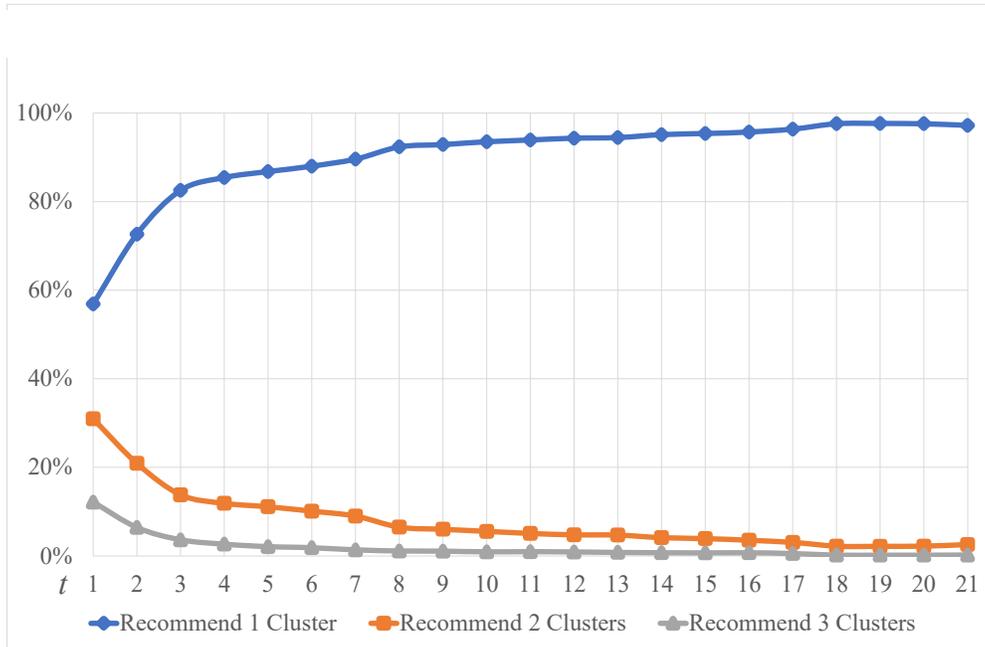}
\end{figure}

Figure \ref{fig:reg_3bg_var} depicts the relative frequencies of
three recommendation cases. The top line (blue) depicts the number
of times the recommendation system opts to serve all recommendations
from a single cluster (e.g., three vehicles from cluster 2). The middle
line (orange) depicts the number of times vehicles from two clusters
are recommended (e.g., one vehicle from cluster 1 and two vehicles
from cluster 2). The bottom line (gray) depicts the number of times
all three vehicles originated from different clusters (e.g., one vehicle
from clusters 1, 2, and 7 each).

Figure \ref{fig:reg_3bg_var} shows that recommendations become more
concentrated as consumers progress in their search journeys. From
around a 60\% likelihood in the beginning, by period 20 the chances
that all recommendations belong to the same cluster have practically
increased to 100\%. This striking pattern is consistent with the recommendation
system learning over time: As consumers reveal more information through
their search actions, the recommendation system hones down on their
``types'' and recommends vehicles from the most appropriate cluster,
transitioning from an exploration to an exploitation regime.

\subsection{Recommendation System Value Decomposition\label{subsec:Rec.-System-Value}}

In this section, we aim to understand how different recommendation
system features and data drive value creation for the company. Our
approach is based on ``comparative statics'': We turn different
features on and off to assess their effects on the performance of
the recommendation system, as measured by expected profit. This approach
is in contrast with the majority of the literature on recommendation
systems, which tends to focus on model performance rather than ask
why certain performance gains arise.\\\\
\textbf{Recommendation Matrices. }We start out by comparing four scenarios.
Scenario 1 corresponds to the performance of the status quo recommendation
policy depicted in Table \ref{tab:status-quo-rec8}; Scenario 2 reoptimizes
the stationary status quo probabilities (i.e., it optimizes the status
quo recommendation system); Scenario 3 considers a recommendation
system based on time-dependent recommendation matrices. In this case,
the consumer is exposed to different recommendation policies, drawn
from different recommendation matrices, as she proceeds through the
website. 

\begin{figure}[H]
\caption{Performance Benchmarks and Rec. Matrix Optimization\label{fig:Benchmark1}}

\centering{}%
\begin{tabular}{c}
\includegraphics[width=0.7\paperwidth]{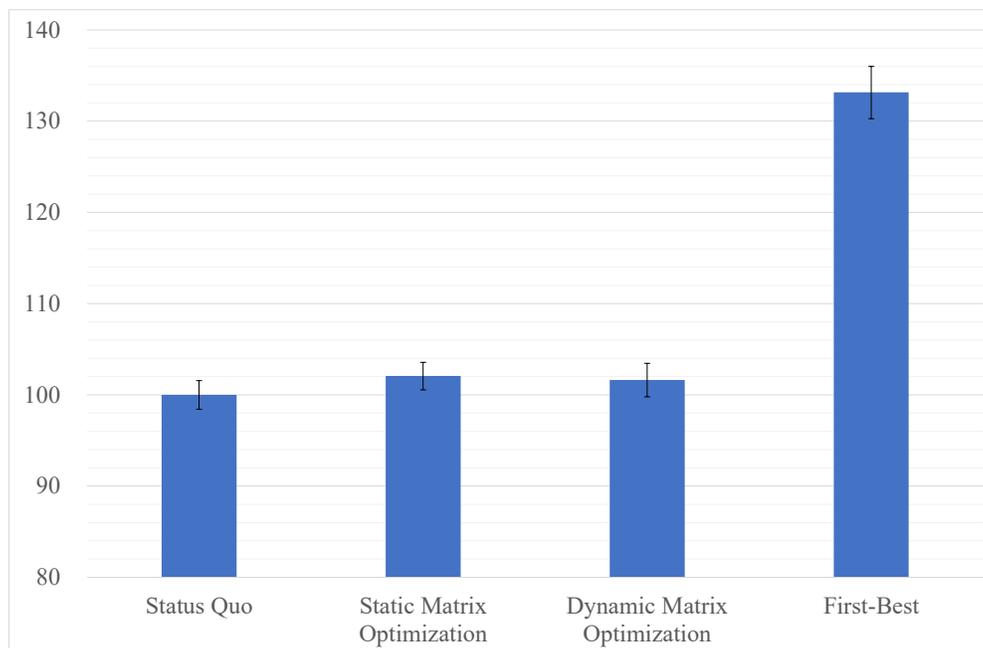}\tabularnewline
\begin{minipage}[t]{16cm}%
Note: Values of expected profit are presented in relation to scenario
``Status Quo'', which has been normalized to 100. Bands represent
one standard deviation of estimated profit.%
\end{minipage}\tabularnewline
\end{tabular}
\end{figure}

Scenario 4 corresponds to the first-best scenario. The results are
presented in Figure \ref{fig:Benchmark1}.\footnote{All results from the counterfactual analyses are shown simultaneously
in Appendix \ref{sec:Counterfactual-Analyses-1}, Table \ref{tab:Expected-Profits-and-Std-Dev}.}

We start out by normalizing the expected profit of the Status Quo
scenario (current recommendation system) to 100. As such, all other
scenarios can be understood as relative changes to the status quo
case (i.e., an expected profit of 110 means the expected profit is
10\% higher than the status quo case). We do this for ease of comparison
and for confidentiality purposes.

Analysis of Figure \ref{fig:Benchmark1} reveals that the seller's
recommendation system does extremely well when compared to the second
scenario, in which we reoptimize the recommendation matrix (profit
increases to 102.1, and difference is not statistically significant).
The third scenario, in which we allow for different recommendation
matrices for each time period (i.e., the order of vehicles viewed)
also does not produce a meaningfully different profit. In fact, we
find that expected profit is equal to 101.6, which is above the status
quo scenario, but below the static recommendation matrix case.\footnote{Optimizing recommendation matrices can be a more demanding task than
finding the ``unconstrained'' first-best recommendation policy.
We describe the approximation methods employed in Appendix \ref{sec:Optimization-of-the-rec-matrices}.} Finally, by conducting bootstrap sampling of the consumer policy,
we find that the expected profits of the three regimes are not statistically
significantly different ($p$-values marginally above 0.05).

The last column of Figure \ref{fig:Benchmark1} depicts expected profit
when full maximization is employed (i.e., recommendation policy is
optimized for each possible consumer state). In this case, we find
that the highest expected profit the company should expect is of 133.2.
Thus, we estimate that there is significant margin to optimize the
recommendation system, although none of the matrix methods is able
to capture a significant fraction of that potential.\newpage{}

\begin{figure}[H]
\caption{Performance of Different Data Usage Scenarios\label{fig:Benchmark2} }

\centering{}%
\begin{tabular}{c}
\includegraphics[width=0.7\paperwidth]{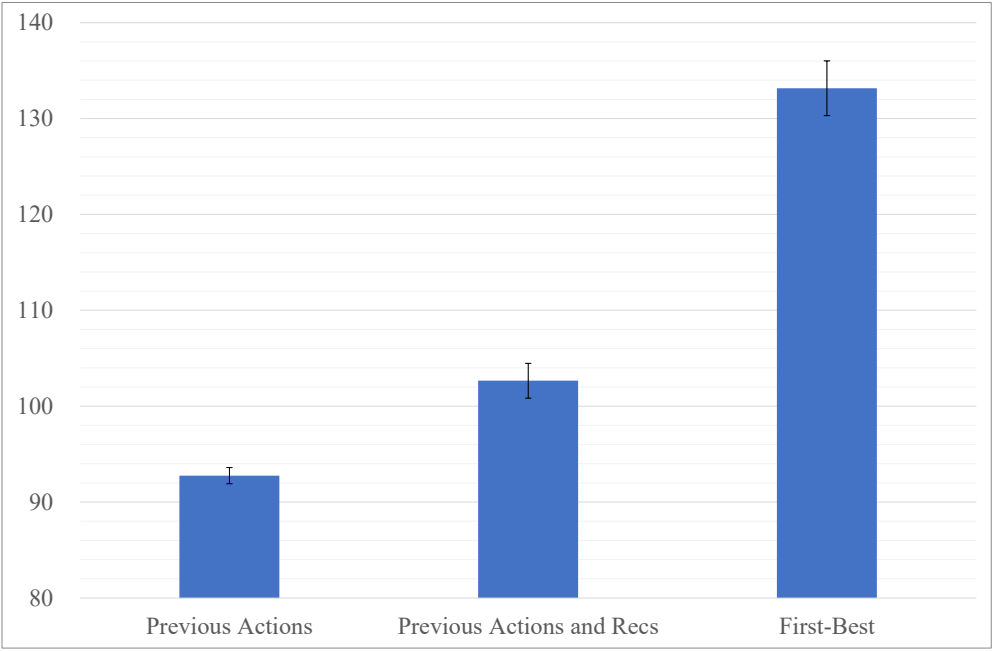}\tabularnewline
\begin{minipage}[t]{16cm}%
Note: Values of expected profit are presented in relation to scenario
``Status Quo'', which has been normalized to 100. Bands represent
one standard deviation of estimated profit.%
\end{minipage}\tabularnewline
\end{tabular}
\end{figure}

\textbf{Value of History.} Figure \ref{fig:Benchmark2} presents how
the use of different data affect the performance of the recommendation
system. In all scenarios below, we calculate the first-best recommendation
system, while introducing data constraints. In the first case, we
optimize the recommendation system only based on the consumer's previous
browsing actions, ignoring the current vehicle being viewed. In the
second scenario we introduce previous vehicle recommendations to the
optimization program. Finally, the difference between the second scenario
and the first-best case is the introduction of the vehicle presently
being viewed by the consumer into the recommendation system optimization.

First, we find that optimizing the recommendation system based solely
on consumers' previous actions does significantly worse than the status
quo (expected profit=93). This result suggests that consumers' past
actions are not necessarily very indicative of preferences, since
history does not lend itself to producing a large portion of future
recommendation value. This result is in contrast with the view that
past behavioral data is fundamental to predict future behavior. In
our case, actions taken before the present one produce little value.
It is possible that this insight applies mostly to search contexts,
in which alternatives that have been passed on are likely to reveal
more about what consumers \emph{are not} interested in, rather than
what consumers \emph{are} interested in.

\begin{figure}[H]
\caption{Performance of Different Rec. System Features\label{fig:Benchmark3}}

\centering{}%
\begin{tabular}{c}
\includegraphics[width=0.7\paperwidth]{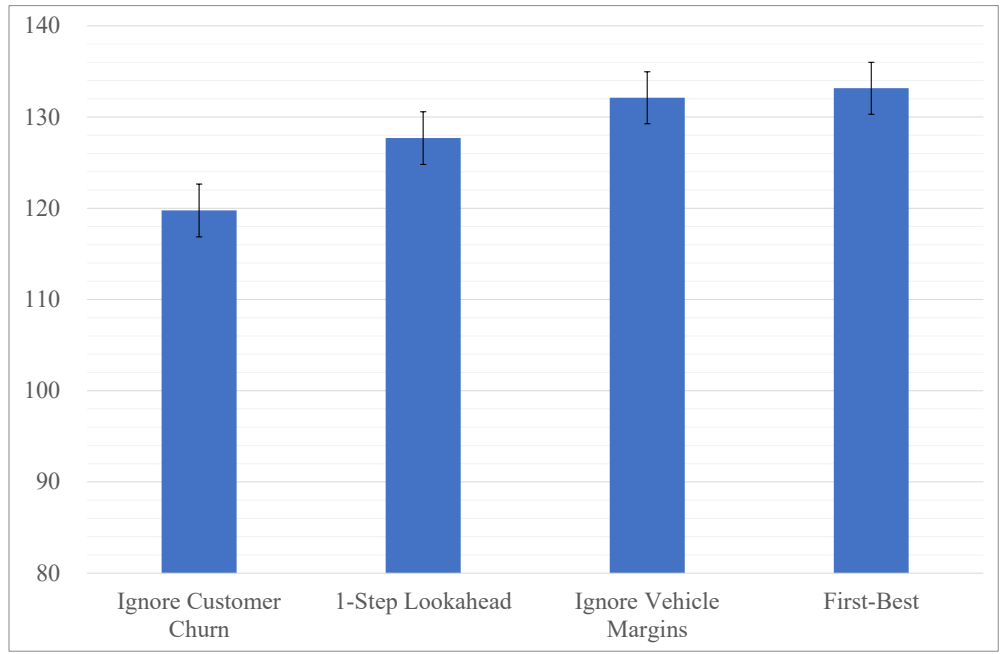}\tabularnewline
\begin{minipage}[t]{16cm}%
Note: Values of expected profit are presented in relation to scenario
``Status Quo'', which has been normalized to 100. Bands represent
one standard deviation of estimated profit.%
\end{minipage}\tabularnewline
\end{tabular}
\end{figure}

Including recommendations shown to the consumer in the past increases
the recommendation system performance significantly above the status
quo, to 102.7 (p\textless 0.01). More noticeable is the difference
between the second and third columns, which is explained by the introduction
of the current vehicle being viewed in to the recommendation system
optimization. In this case (the first-best), expected profit increases
from 102.7 to 133. Ensuring that the current option being browsed
feeds into the recommendation system seems to be the most important
behavioral data piece, as it generates the highest increase in expected
profit. Together with the previous results, this analysis proposes
a strong complementary effect between the vehicle currently being
viewed and the remaining data available to the recommendation system.\\\\
\textbf{Recommendation Goals. }Next, we describe how different features
of the recommendation system drive expected profit (Figure \ref{fig:Benchmark3}).
In this analysis, the recommendation system is optimized based on
the full dataset. However, the objective function is altered in each
scenario.

The first bar corresponds to the case where the recommendation system
assigns a low probability to the event of consumers abandoning the
website (outside option). In this case, the probability of selecting
the outside option is reduced to zero during model optimization, while
the probabilities of the remaining actions are increased in proportion.
In other words, in this case the recommendation system acts as if
the consumer will buy for sure. The analysis reveals that ignoring
the possibility of consumer churn limits expected profit significantly,
to 120, as compared with the first-best scenario of 133.

The second column turns off forward-looking behavior by the recommendation
system. In this case, the recommendation system acts so as to optimize
the next period's payoff, but ignores any subsequent ones. The underlying
goal of this analysis is to understand the value of ``strategic behavior''
in recommendations. For example, when the recommendation system is
perfectly forward looking, it can suggest vehicles that may not appear
optimal immediately -- and may not be even bought -- but may lead
consumers to explore other vehicles that will later increase the probability
of conversion and margins realized. Our analysis estimates that ignoring
forward-looking behavior decreases expected profits from 133 to 127.7.
The difference is significant both in magnitude (5\% of status quo
profits) as well as statistically.

We conclude by conducting a counterfactual scenario in which the recommendation
system ignores the fact that different vehicles yield different profit
margins for the firm. For example, in the first-best scenario the
algorithm may recommend products with higher margins, on average,
whereas in the ``Ignore Business Goals'' scenario it cares only
about the probability of conversion. Perhaps surprisingly, expected
profit changes very little (133 to 132.1) when product margins are
ignored, and the difference is not statistically significant.

As a whole, these results indicate that one of the the most important
tasks for a recommendation system is to keep the consumer engaged
and away from churning. In contrast, forward-looking recommendations
generate relatively modest benefits, and taking product margins into
consideration affect profits only slightly.

\section{Final Remarks\label{sec:Conclusion}}

Recommendation systems are extremely popular among sellers. So far,
the emphasis on improving such systems is through the acquisition
of larger and better customer data, and the development of hybrid
recommendation methods. This trend has left small businesses, sellers
constrained by the very nature of their transactions, and others abiding
by strict privacy regulations, out of the conversation. In this paper
we develop a recommendation system that addresses the ``cold start''
problem, a blocking factor for such actors.

The model is inspired in a fundamental insight from the search literature:
Consumers do not hold static views on products, but they update them
as they acquire information. As a result, it is important to take
the consumer's state into account when calculating recommendations.
For example, our model takes into account that a given recommendation
may affect subsequent consumer search, leading to a different set
of future search and conversion actions. The model relies on standard
Bellman equation formulation and dynamic programming techniques (state
aggregation and value function interpolation). It takes advantage
of the institutional features of our setting, which allow us to recover
the causal effect of recommendations from revealed-preference data.
Our analysis suggests that the model transitions from exploration
to exploitation, as it suggests more similar alternatives along the
consumer search journey.

Finally, we investigate the resulting `black box' by conducting a
series of counterfactual analyses. We find that it is the joint effect
of past behavioral and recommendation data with the current vehicle
being browsed the allows the model to drive the most value to the
seller. On their own, each of these data components account for small
improvements in ex-ante profits. The final set of counterfactuals
reveals that managing consumer churn is a primary concern for the
recommendation model. In addition, the recommendation system benefits
from thinking not only about the next consumer decision, but the whole
path until conversion.

Currently, most recommendation systems rely on a hybridization of
collaborative filtering and content-based filtering. Despite their
popularity, these systems tend to rely on an implicit estimate of
each product's worth to each consumer. It would be interesting to
investigate the extent to which our approach can be combined with
the traditional ones, so as to alleviate the cold start problem while
incorporating the sequential nature of consumer search into product
recommendations.

\newpage{}

\appendix

\section*{\emph{Appendix}}

\appendix

\section{The Consumer's Dynamic Search Problem\label{Appendix:Model}}

\subsection{Dynamic Problem}

Consumer $i$ derives utility $u_{ijt}$ from consuming alternative
$j\in\{1,...,J\}$. The utility from the outside good $j=0$ is normalized
to $u_{0}=\epsilon_{0}$. Consumers do not know $u_{ijt}$ a priori,
but hold prior beliefs about the distribution of $u_{ijt}$, $F(u_{ijt}|\Theta_{it}^{'}$).
That prior distribution depends on the information state, $\Theta_{it}^{'}$
at every period $t=0,..,T$. We omit the individual-level subscripts
from here on, for simplicity.

In every period, the consumer decides between (i) searching an alternative
$j\in\{1,...,J\}$ (extensive margin); (ii) terminating search and
purchase an alternative $(j=0,...,J)$, where $j=0$ is the outside
good.

Performing a search action has a cost of $c$.\footnote{Notice that our approach also allows the search cost to be time dependent.}
The dynamic problem faced by the consumer can be described by the
solution to the following Bellman equation

\begin{equation}
V\left(\Theta_{t}^{'}\right)=\max\left\{ \underset{Conversion\,Actions}{\underbrace{\left.E(u_{ijt}|\Theta_{t}^{'})\right|_{j=0,...,J}}},\underset{Search\,Actions}{\underbrace{\left.-c_{i}+E(V(\Theta_{i,t+1}|\Theta_{t}^{'})\right|_{j=J+1,...2J+1}}}\right\} \label{eq:cons_dyn_prob}
\end{equation}

In each period, the consumer can make one of $2J+1$ decisions. As
researchers, we observe the decisions related with search and conversion.
The solution to the problem above, $y_{it}$, as a function of state
space $\Theta_{it}^{'}$, is the relationship the researchers observe
and estimate from the data.

\subsection{A Tale of Two Dynamic Problems}

We now clarify how the consumers' dynamic problem intertwines with
the recommendation system's. Table \ref{tab:Sequence-of-Consumer}
exemplifies the decisions and state transitions of both parties.

\begin{table}[H]
\caption{Sequence of Consumer and Recommendation System Decisions\label{tab:Sequence-of-Consumer}}

\centering{}{\footnotesize{}}%
\begin{tabular}{>{\centering}m{2cm}>{\centering}m{3cm}>{\centering}m{2cm}|>{\centering}m{0.5cm}|>{\centering}m{4cm}>{\centering}m{3cm}}
 &  & \multicolumn{1}{>{\centering}m{2cm}}{} & \multicolumn{1}{>{\centering}m{0.5cm}}{} &  & \tabularnewline
\hline 
\hline 
{\footnotesize{}$t$} & {\footnotesize{}User State $\left(\Theta_{t}^{'}\right)$} & \multicolumn{1}{>{\centering}m{2cm}}{{\footnotesize{}User Action}} & \multicolumn{1}{>{\centering}m{0.5cm}}{} & {\footnotesize{}Recommendation System's State $\left(\Theta_{t}\right)$} & {\footnotesize{}Rec. System Action}\tabularnewline
\hline 
{\footnotesize{}$t=1$} & {\footnotesize{}}%
\begin{minipage}[t]{2cm}%
{\footnotesize{}
\begin{align*}
\Theta_{1}^{'}:t= & 1\\
a_{0}= & \emptyset\\
A_{0}= & \emptyset\\
R_{1}= & \emptyset
\end{align*}
}%
\end{minipage}{\footnotesize{}}\\
\textcolor{white}{\footnotesize{}.} & {\footnotesize{}$a_{1}\left(\Theta_{1}^{'}\right)=3$} & $\rightarrow$ & {\footnotesize{}}%
\begin{minipage}[t]{2cm}%
{\footnotesize{}
\begin{align*}
\Theta_{1}:t= & 1\\
a_{1}= & 3\\
A_{1}= & \emptyset\\
R_{1}= & \emptyset
\end{align*}
}%
\end{minipage}{\footnotesize{}}\\
\textcolor{white}{\footnotesize{}.} & {\footnotesize{}$r_{2}\left(\Theta_{1}\right)=1$}\tabularnewline
\hline 
{\footnotesize{}$t=2$} & {\footnotesize{}}%
\begin{minipage}[t]{2cm}%
{\footnotesize{}
\begin{align*}
\Theta_{2}^{'}:t= & 2\\
a_{1}= & 3\\
A_{1}= & \emptyset\\
R_{2}= & \{1,0,0\}
\end{align*}
}%
\end{minipage}{\footnotesize{}}\\
\textcolor{white}{\footnotesize{}.} & {\footnotesize{}$a_{2}\left(\Theta_{2}^{'}\right)=1$} & $\rightarrow$ & {\footnotesize{}}%
\begin{minipage}[t]{2cm}%
{\footnotesize{}
\begin{align*}
\Theta_{2}:t= & 2\\
a_{2}= & 1\\
A_{2}= & \{0,0,1\}\\
R_{2}= & \{1,0,0\}
\end{align*}
}%
\end{minipage}{\footnotesize{}}\\
\textcolor{white}{\footnotesize{}.} & {\footnotesize{}$r_{3}\left(\Theta_{2}\right)=3$}\tabularnewline
\hline 
$t=3$ & {\footnotesize{}}%
\begin{minipage}[t]{2cm}%
{\footnotesize{}
\begin{align*}
\Theta_{3}^{'}:t= & 3\\
a_{2}= & 1\\
A_{2}= & \{0,0,1\}\\
R_{3}= & \{\nicefrac{1}{2},0,\nicefrac{1}{2}\}
\end{align*}
}%
\end{minipage}{\footnotesize{}}\\
\textcolor{white}{\footnotesize{}.} & {\footnotesize{}$a_{3}\left(\Theta_{3}^{'}\right)=2$} & $\rightarrow$ & ... & \tabularnewline
\hline 
\end{tabular}{\footnotesize\par}
\end{table}

At $t=1$, the user arrives to the seller's homepage and selects a
vehicle, $a_{1}\left(\Theta_{1}^{'}\right)=3$. At this point, her
state is mostly empty, since no history exists. The recommendation
system then proposes a vehicle from cluster $1$, $r_{2}\left(\Theta_{1}\right)=1$
(assume a single recommendation for clarity).\footnote{Also, since the first consumer action is made without recommendations
(a click on the homepage), $r_{1}$ is not defined.} Notice that the states used by each party to inform decision-making
are different. Specifically, the recommendation is calculated upon
the consumer's last decision, and the consumer's decision is based
on the recommendation system's last recommendation.

At $t=2$, the consumer's state comprises the time period, the last
consumer action $\left(a_{1}=3\right)$, the actions before that $\left(A_{1}=\emptyset\right)$
and the previous recommendations, $R_{2}=R_{1}\cup r_{2}=\{1,0,0\}$.
Based on these data, the consumer opts for decision $a_{2}\left(\Theta_{2}^{'}\right)=1$.
In response, the recommendation system updates its state to $\left\{ t=2,a_{2}=1,A_{2}=\{0,0,1\},R_{2}=\{1,0,0\}\right\} $.
At this point, $a_{2}$ captures the vehicle that the consumer wants
to browse now, whereas $A_{2}=\{0,0,1\}$ holds the consumer's first
browsing decision, i.e., $A_{2}=A_{1}\cup a_{1}$. Finally, the recommendation
system opts to recommend a vehicle from cluster 3, and we move to
period $3$. The consumer's state updating proceeds in the same fashion,
with state variable $R_{3}$ being calculated via relative frequencies
of past recommendations, $R_{3}=\frac{1}{2}\left(R_{1}+R_{2}\right)=\{\nicefrac{1}{2},0,\nicefrac{1}{2}\}$.

\section{Optimization of the Static and Dynamic Matrices\label{sec:Optimization-of-the-rec-matrices}}

Consider the value function faced by the recommendation system once
again:
\begin{align}
V\left(t,a_{t},A_{t},R_{t}\right) & =\max_{\varphi\left(\cdot\right)}\,\pi\left(t,a_{t},A_{t},R_{t}\right)+E\left[\left.V\left(t+1,a_{t+1},A_{t+1},R_{t+1}\right)\right|a_{t},A_{t},R_{t},\varphi\left(\cdot\right)\right]\label{eq:value_fn_fb-2}
\end{align}

For counterfactuals ``Static Matrix Optimization'' and ``Dynamic
Matrix Optimization'', problem \eqref{eq:value_fn_fb-2} is solved
under the constraint that $\varphi\left(\cdot\right)$ is an $8\text{\ensuremath{\times8}}$
recommendation matrix of probabilities, which is constant or can change
over time, depending on the scenario of interest. This optimization
has many control variables (in the dynamic case, this yields 1,232
variables) and is infeasible through direct optimization methods.
Our optimization approach is to solve problem \eqref{eq:value_fn_fb-2}
once, based on the original dataset, and then, for each bootstrap
sample, approximate problem \eqref{eq:value_fn_fb-2} quadratically
via Taylor series expansion, and calculate the resulting maximizer
analytically.

Let the original policy function, estimated from the original dataset,
induce the following dynamic problem for the recommendation system:
\[
V\left(t,a_{t},A_{t},R_{t}\right)=\max_{\varphi\left(\cdot\right)}\,\pi\left(t,a_{t},A_{t},R_{t}\right)+E_{0}\left[\left.V\left(t+1,a_{t+1},A_{t+1},R_{t+1}\right)\right|a_{t},A_{t},R_{t},\varphi\left(\cdot\right)\right]
\]
where $E_{0}$ incorporates the consumer's policy function recovered
via machined learning estimation. We first define the following non-optimized
function 
\begin{equation}
W\left(\varphi,f\left(\cdot\right)\right):=\pi\left(t,a_{t},A_{t},R_{t}\right)+E\left[\left.V\left(t+1,a_{t+1},A_{t+1},R_{t+1}\right)\right|a_{t},A_{t},R_{t},\varphi\right]\label{eq:function_W}
\end{equation}
Note that function $W$ is a functional of the conditional probability
mass function 
\begin{equation}
f(\cdot)\text{=}f\left(\left.a_{t+1}\right|t,a_{t},A_{t},R_{t},\varphi\right)\label{eq:pmf1}
\end{equation}
which is the estimated object summarizing the consumer policy function.
Denote the analogue object recovered from the original dataset by
$f_{0}$. Through classical numerical solvers, we can calculate 
\begin{equation}
\varphi_{0}^{*}=\text{argmax}_{\varphi}\,W\left(\varphi,f_{0}\right)\label{eq:W2}
\end{equation}
where $\varphi_{0}^{*}$ is a large matrix. Taylor expansion yields
the following quadratic approximation for function $W$ evaluated
at bootstrap policy sample $b$:
\begin{equation}
\overset{\sim}{W}\left(\varphi,f_{b}\right)=W\left(\varphi_{0}^{*},f_{b}\right)+D'.\left(\varphi-\varphi_{0}^{*}\right)+\frac{1}{2}\left(\varphi-\varphi_{0}^{*}\right)^{T}.H.\left(\varphi-\varphi_{0}^{*}\right)\label{eq:Taylor}
\end{equation}
where $D=\frac{\partial}{\partial\varphi}W\left(\varphi,f_{b}\right)$
is a matrix of first partial derivatives of $W$, and $H=\frac{\partial^{2}}{\partial\varphi^{2}}W\left(\varphi,f_{b}\right)$
is the (symmetric) Hessian matrix of $W$, both evaluated at $\varphi=\varphi_{0}^{*}$.
Taking the first-order condition of the right-hand side w.r.t. $\text{\ensuremath{\varphi}}$
yields:
\begin{align*}
\frac{\partial}{\partial\varphi}\overset{\sim}{W}\left(\varphi,f_{b}\right) & =0\\
\Leftrightarrow0+D'+\frac{1}{2}\left(2\varphi'H-2\varphi_{0}^{*'}H\right) & =0\\
\varphi'H & =-D'+\varphi_{0}^{*'}H\\
\varphi' & =\varphi_{0}^{*'}-D'H^{-1}\\
\varphi & =\varphi_{0}^{*}-H^{-1}D
\end{align*}
where we have taken advantage of the fact that $H$ is a symmetric
matrix. For each sample $f_{b}$, this procedure yields a matrix recommendation
$\varphi_{b}$, at the cost of evaluating the first and second matrices
of derivatives of $W$ at each estimated bootstrap policy function
$f_{b}$, calculated through finite differences. The optimization
of the static matrix scenario across bootstrap samples took the longest
to calculate, and was parallelized within and across computers (takes
about 22 days on a top-of-the-line 8-core pc).

\section{Counterfactual Analyses\label{sec:Counterfactual-Analyses-1}}
\begin{center}
\begin{table}[H]
\caption{Expected Profits and their Standard Deviations for each Counterfactual
Scenario\label{tab:Expected-Profits-and-Std-Dev}}

\begin{centering}
\begin{tabular}{lcc}
 &  & \tabularnewline
\hline 
\hline 
Scenario & Expected Profit & Std. Deviation\tabularnewline
\hline 
Status Quo & 100.0 & 1.57\tabularnewline
Static Matrix Optimization & 102.1 & 1.51\tabularnewline
Dynamic Matrix Optimization & 101.6 & 1.84\tabularnewline
Previous Actions Only & 92.8 & 0.84\tabularnewline
Previous Actions and Recommendations & 102.7 & 1.81\tabularnewline
Ignore Vehicle Margins & 132.1 & 2.85\tabularnewline
Ignore Customer Churn & 119.8 & 2.85\tabularnewline
1-Step Look-ahead & 127.7 & 2.90\tabularnewline
First-Best Policy & 133.2 & 2.89\tabularnewline
\hline 
\multicolumn{3}{c}{%
\begin{minipage}[t]{12cm}%
Note: Standard deviations calculated based on 50 bootstrap replications
of the consumer policy.%
\end{minipage}}\tabularnewline
\end{tabular}
\par\end{centering}
\end{table}
\par\end{center}

\newpage{}

\bibliographystyle{bibstyle}
\bibliography{writeup_oct_2020}

\end{document}